\documentclass[10pt,twocolumn,letterpaper]{article}

\usepackage{cvpr}              
\usepackage[accsupp]{axessibility} 

\usepackage{graphicx}
\usepackage{amsmath, amsthm}
\usepackage{amssymb}
\usepackage{bm}
\usepackage{booktabs}
\usepackage{framed,multirow}
\usepackage{algorithm, algpseudocode}
\usepackage{varwidth}
\usepackage{thmtools, thm-restate}
\def\NoNumber#1{{\def\alglinenumber##1{}\State #1}\addtocounter{ALG@line}{-1}}
\usepackage{adjustbox}
\usepackage{tabularx}
\usepackage{makecell}

\newcommand{\rev}[1] {\textcolor{red}{#1}} 

\newtheorem{theorem}{\bf{Theorem}}
\newtheorem{lemma}{\bf{Lemma}}

\newtheorem{corollary}{Corollary}
\newtheorem{definition}{\bf{Definition}}

\def\code#1{\texttt{#1}}

\long\def\comment#1{} 

\newcommand{\xmath}[1] {\ensuremath{#1}\xspace}
\newcommand{\blmath}[1] {\xmath{\bm{#1}}}
\newcommand{\A}{\blmath{A}}

\newcommand{\D}{\blmath{D}}

\newcommand{\F}{\blmath{F}}
\newcommand{\I}{\blmath{I}}

\renewcommand{\P}{\blmath{P}} 
\renewcommand{\F}{\blmath{F}} 

\newcommand{\x}{\mathbf{x}}

\newcommand{\y}{\blmath{y}}
\newcommand{\z}{\blmath{z}}
\newcommand{\f}{\blmath{f}}

\newcommand{\w}{\blmath{w}}
\newcommand{\s}{\blmath{s}}

\newcommand{\0}{\blmath{0}}

\newcommand{\norm}[1] {\xmath{\left\| #1 \right\|}}

\newcommand{\Ab}{{\blmath A}}

\newcommand{\Ib}{{\blmath I}}

\newcommand{\Pb}{{\blmath P}}

\newcommand{\bb}{{\blmath b}}

\newcommand{\fb}{{\blmath f}}

\newcommand{\hb}{{\blmath h}}

\renewcommand{\sb}{{\blmath s}}

\newcommand{\wb}{{\blmath w}}
\newcommand{\xb}{{\blmath x}}
\newcommand{\yb}{{\blmath y}}
\newcommand{\zb}{{\blmath z}}

\newcommand{\Nc}{\mathcal{N}}

\def\code#1{\texttt{#1}}

\newcommand{\Rd}{{\mathbb R}}

\newcommand{\Nd}{{\mathbb N}}

\newcommand{\thetab}{{\boldsymbol {\theta}}}

\newcommand{\Ed}{{{\mathbb E}}}

\newcommand{\beq}{\begin{equation}}
\newcommand{\eeq}{\end{equation}}
\newcommand{\beqa}{\begin{eqnarray}}
\newcommand{\eeqa}{\end{eqnarray}}

\renewcommand{\x}{\xb}
\usepackage[pagebackref,breaklinks,colorlinks]{hyperref}

\usepackage[capitalize]{cleveref}
\crefname{section}{Sec.}{Secs.}
\Crefname{section}{Section}{Sections}
\Crefname{table}{Table}{Tables}
\crefname{table}{Tab.}{Tabs.}

\newcommand\blfootnote[1]{%
  \begingroup
  \renewcommand\thefootnote{}\footnote{#1}%
  \addtocounter{footnote}{-1}%
  \endgroup
}
\def\cvprPaperID{11776} 
\def\confName{CVPR}
\def\confYear{2022}

\begin{document}

\title{Come-Closer-Diffuse-Faster: Accelerating Conditional Diffusion Models for Inverse Problems through Stochastic Contraction}
\vspace{-0.5cm}
\author{Hyungjin Chung$^1$ \quad\quad Byeongsu Sim$^2$ \quad\quad Jong Chul Ye$^{1,2,3}$\\
$^1$Bio and Brain Engineering, $^2$Mathematical Sciences, 
 $^3$Kim Jaechul Graduate School of AI\\
Korea Advanced Institute of Science and Technology (KAIST),  Daejeon, Korea\\
{\tt\small \{hj.chung, byeongsu.s, jong.ye\}@kaist.ac.kr}
}
\twocolumn[{%
\renewcommand\twocolumn[1][]{#1}%
\maketitle
\vspace{-0.5cm}
\begin{center}
\vspace{-0.7cm}
\includegraphics[width=0.87\linewidth]{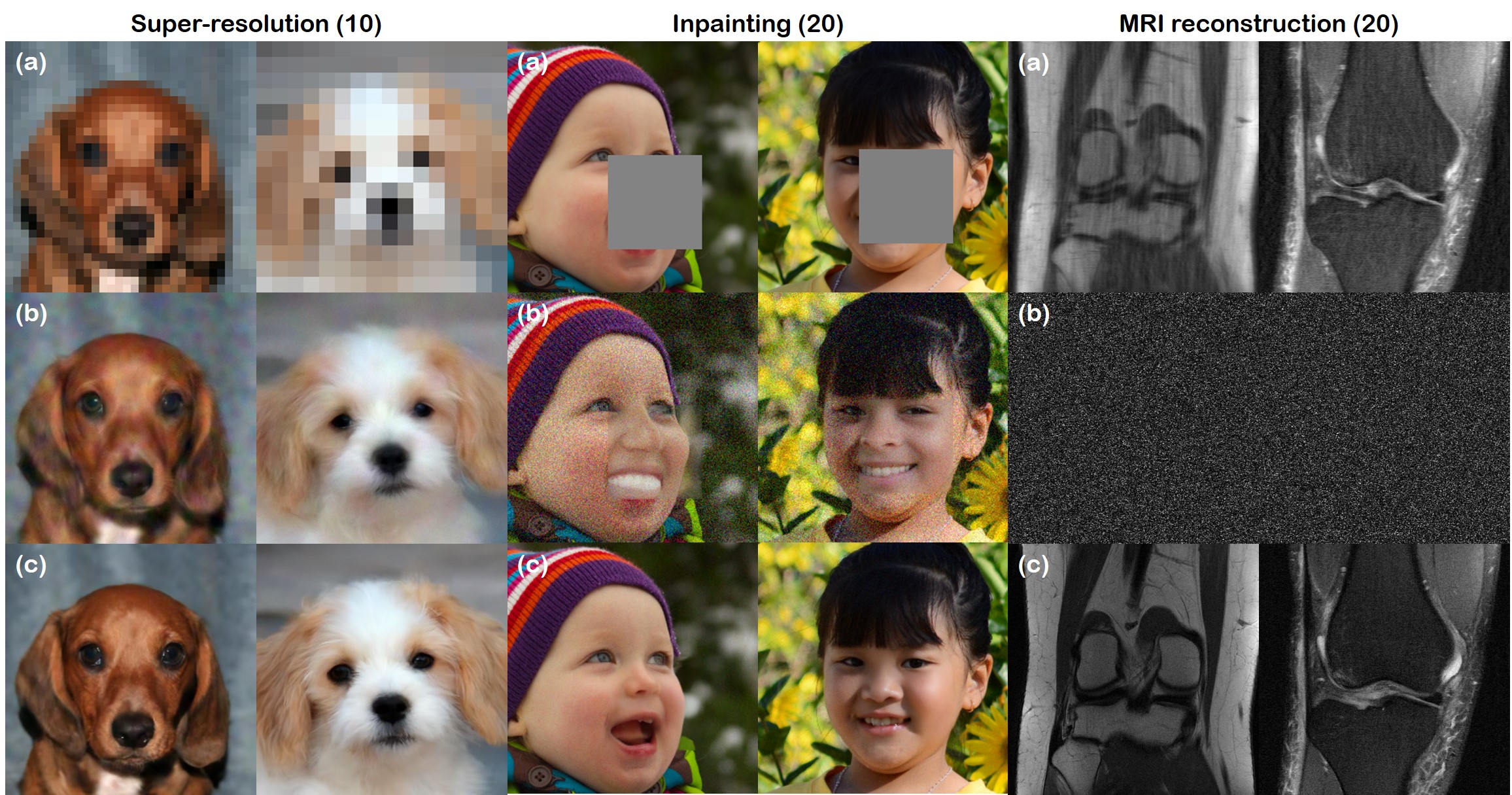}
\vspace{-0.2cm}
\captionof{figure}{Reconstruction results of three different tasks - super-resolution, inpainting, and MRI reconstruction. Numbers in parenthesis indicate the iteration numbers  for reverse diffusion. Proposed method is compared with canonical conditional diffusion models for each task. {(a) Corrupted measurement, (b) ILVR~\cite{choi2021ilvr}, score-SDE~\cite{song2020score}, and {score-MRI}~\cite{chung2021score}, respectively, for each task. (c) Proposed method.}}
   \label{fig:concept}
\end{center}
}]

\blfootnote{This work was supported by Institute of Information \& communications Technology Planning \& Evaluation (IITP) grant funded by the Korea government(MSIT) (No.2019-0-00075, Artificial Intelligence Graduate School Program(KAIST)), and by National Research Foundation(NRF) of Korea grant NRF-2021M3I1A1097938}
\begin{abstract}
Diffusion models have recently attained significant interest within the community owing to their strong performance as generative models. Furthermore, its application to inverse problems have demonstrated state-of-the-art performance. Unfortunately, diffusion models have a critical downside - they are inherently slow to sample from, needing few thousand steps of iteration to generate images from pure Gaussian noise.  In this work, we show that starting from Gaussian noise is unnecessary. Instead, starting from a single forward diffusion with better initialization significantly reduces the number of sampling steps in the reverse conditional diffusion. This phenomenon is formally explained by the contraction theory of the stochastic difference equations like our conditional  diffusion strategy - the alternating applications of reverse diffusion followed by a non-expansive data consistency step. The new sampling strategy, dubbed Come-Closer-Diffuse-Faster (CCDF), also reveals a new insight on how the existing feed-forward neural network approaches for inverse problems can be synergistically combined with the  diffusion models. Experimental results with super-resolution, image inpainting, and compressed sensing MRI demonstrate that our method can achieve state-of-the-art reconstruction performance at significantly reduced sampling steps.
\end{abstract}

\vspace{-0.3cm}
\section{Introduction}
\label{sec:intro}

\begin{figure*}[!hbt]
    \centering\includegraphics[width=17cm]{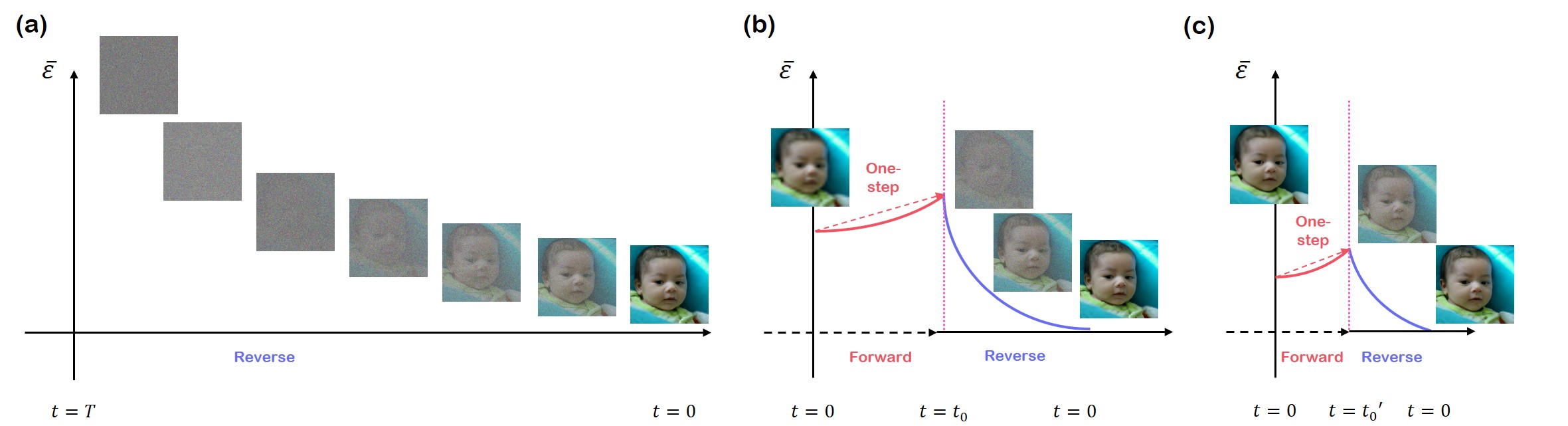}
    \caption{{Plot of average error $\bar{\varepsilon}$ vs. time $t$, using different approaches. (a) Conditional diffusion starts from Gaussian noise $\x(t)$ and uses full reverse diffusion. (b) CCDF with vanilla initialization: Corrupted data is forward-diffused with a single step up to $t = t_0$, and reverse diffused. (c) CCDF with NN initialization:  Initialization with reconstruction from pre-trained NN lets us use much smaller timestep $t = t'_0 < t_0$, and hence faster reverse diffusion.}}
	\label{fig:CCDF_concept}
\end{figure*}

Denoising diffusion models~\cite{sohl2015deep,ho2020denoising,dhariwal2021diffusion, kingma2021variational} and score-based models~\cite{song2019generative,song2020improved,song2020score} are new trending classes of generative models, which have recently drawn signficant attention amongst the community due to their state-of-the-art performance. Although inspired differently, both classes share very similar aspects, and can be cast as variants of each other~\cite{song2020score,huang2021variational,kingma2021variational}, thus they are often called {\em diffusion models}. 

In the forward diffusion process, a sampled data point $\x$ at time $t = 0$ is perturbed gradually with Gaussian noise until $t = T$, arriving approximately at spherical Gaussian distribution, which is easy to sample from. In the reverse diffusion process, starting from the sampled noise at $t = T$, one uses the trained score function to gradually denoise the data up to $t = 0$, arriving at a high quality data sample. 

Interestingly, diffusion models can go beyond unconditional image synthesis, and have been applied to conditional image generation, including super-resolution~\cite{saharia2021image, choi2021ilvr, li2021srdiff}, inpainting~\cite{song2019generative, song2020score}, MRI reconstruction~\cite{chung2021score, song2022solving, jalal2021robust}, image translation~\cite{meng2021sdedit, choi2021ilvr, sasaki2021unit}, and so on. One line of works re-design the diffusion model specifically suitable for the task at hand, thereby achieving remarkable performance on the given task~\cite{saharia2021image, li2021srdiff, sasaki2021unit}. However, they compromise flexibility since the model cannot be used on other tasks. Another line of works,
on which we build our method on,
 keep the training procedure intact, and only modify the inference procedure such that one can sample from a conditional distribution~\cite{song2020score, chung2021score, choi2021ilvr, song2022solving, jalal2021robust}. These methods   can be thought of as leveraging the learnt score function as a generative prior of the data distribution, and can be flexibly used across different tasks. 

Unfortunately, a critical drawback of diffusion models is that they are very slow to sample from.  To address this,
for unconditional generative models, many works focused on either constructing deterministic sample paths from the stochastic counterparts~\cite{song2020denoising, song2020score}, searching for the optimal steps to take after the training of the score function~\cite{chen2021wavegrad, watson2021learning}, or by retraining student networks that can take shortcuts via knowledge distillation~\cite{luhman2021knowledge, salimans2022progressive}. Orthogonal and complementary to these prior works, in this work, we focus on accelerating  {\em conditional} diffusion models by studying the contraction property~\cite{pham2009contraction,pham2008analysis,park2021generative} of the reverse diffusion path.

Specifically, our method, which we call {\em Come-Closer-Diffuse-Faster (CCDF)}, first perturbs the initial estimate via forward {diffusion} path up to $t_0<T$, where {$t_0$} denotes the time where the reverse diffusion {starts}.
This forward diffusion comes almost for free, without requiring any passes through the neural network. While the distribution of forward-diffused (noise-added) images increases the estimation errors from the initialization as shown in Fig.~\ref{fig:CCDF_concept}(b), the key idea of the proposed CCDF is that  the reverse conditional diffusion path reduces the error {\em exponentially} fast thanks to the  contraction property of the stochastic difference equation \cite{pham2009contraction,pham2008analysis}.
Therefore, compared to the standard approach that starts the reverse diffusion from Gaussian distribution at $t=T$  (see Fig.~\ref{fig:CCDF_concept}(a)), the total number of the reverse diffusion step to recover a clean images using  CCDF can be significantly reduced.
Furthermore, with better initialization, we prove that the number of reverse sampling can be further reduced as shown in  Fig.~\ref{fig:CCDF_concept}(c). This implies that the existing neural-network (NN) based inverse solution can be synergistically combined with diffusion models to yield accurate and fast reconstruction by providing {a} better initial estimate.

Using extensive experiments across various problems such as super-resolution (SR), inpainting, and MRI reconstruction, we demonstrate that CCDF {can} significantly {accelerate} diffusion based models for inverse problems.

\section{Background}

\subsection{Score-based Diffusion Models}

We will follow the usual construction of continuous diffusion process $\x(t), t\in[0,T]$ with $\x(t)\in \Rd^d$~\cite{song2020score}. Concretely, we want $\x(0) \sim p_0(\x)$, where $p_0 = p_{\text{data}}$, and $\x(T) \sim p_T$, where $p_T$ is a tractable distribution that we can sample from. 
Consider the following It$\hat{\text{o}}$ stochastic differential equation:
\begin{equation}
\label{eq:forward-sde}
    {d\x = \bar\f(\x, t)dt + \bar g(t)d\w,}
\end{equation}
where {$\bar\f: \Rd^d \mapsto \Rd^d$} is the drift coefficient of $\x(t)$, {$\bar g: \Rd \mapsto \Rd$} is the diffusion coefficient coupled with the standard $d$-dimensional Wiener process $\w \in \Rd^d$. By carefully choosing {$\bar\f, \bar g$}, one can achieve spherical Gaussian distribution as $t \to T$. 
In particular, when {$\bar\f(\x, t)$} is an affine function, then the perturbation kernel $p_{0t}(\x(t)|\x(0))$ is always Gaussian, where the parameters can be calculated in closed-form. Hence, perturbing the data with the perturbation kernel $p_{0t}(\x(t)|\x(0))$ comes almost for free, without requiring any passes through the neural network.

For {the} given forward SDE in \eqref{eq:forward-sde}, there exists a reverse-time SDE running backwards~\cite{song2020score, huang2021variational}:
\begin{align}\label{eq:reverse_SDE}
    d\xb &= {[\bar\fb(\xb, t) - \bar g(t)^2 \underbrace{\nabla_\xb \log p_t(\xb)}_{\text{score function}}]dt + \bar g(t) d\bar{\wb} }
\end{align}
where $dt$ is the infinitesimal {\em negative} time step, and $\bar{\wb}$ is the Brownian motion running backwards. 

Interestingly, one can train a neural network to approximate the actual score function via score matching~\cite{song2019generative, song2020score} to estimate $\s_{\theta}(\x, t) \simeq \nabla_\x \log p_t(\x)$, and plug it into \eqref{eq:reverse_SDE} to numerically solve the reverse-SDE~\cite{song2020score}. 
Furthermore, to circumvent technical difficulties, de-noising score matching is typically used where $\nabla_\x \log p_t(\x)$ is replaced with $\nabla_\x \log p_{0t}(\x(t)|\x(0))$. 

\subsection{Discrete Forms of SDEs}

In this paper, we make use of two different SDEs: variance preserving (VP) SDE, and variance exploding (VE) SDE \cite{song2020score}. First, by choosing 
\begin{equation}\label{eq:VP-SDE}
    {\bar\fb(\xb,t) = -\frac{1}{2}\beta(t)\xb,\quad \bar g(t)=\sqrt{\beta(t)},}
\end{equation}
where $0 < \beta(t) < 1$ is a monotonically increasing function of noise scale, 
one achieves the variance preserving (VP)-SDE \cite{ho2020denoising}.
On the other hand, variance exploding (VE) SDEs choose 
\begin{equation}\label{eq:VE-SDE}
{\bar\fb = \0,\quad \bar g = \sqrt{\frac{d[\sigma^2(t)]}{dt}},}
\end{equation}
where $\sigma(t) > 0$ is again a monotonically increasing function, typically chosen to be a geometric series~\cite{song2019generative, song2020score}. 
 
For the discrete diffusion models, we assume we have $N$ discretizations which are linearly distributed across $t \in [0, T]$.
Then, VP-SDE can be seen as the continuous version of DDPM~\cite{song2020score, kingma2021variational}. 
Specifically,  in DDPM, the forward diffusion is performed as
\begin{align}\label{eq:DDPMf}
\x_{i} = \sqrt{\bar{\alpha}_{i}}{\x}_0 + \sqrt{1 - \bar{\alpha}_{i}}\z
\end{align}
where $\z \sim \Nc(\textbf{0}, \I)$ and $\bar{\alpha}_i = \prod_{j=1}^{i-1} \alpha_j$ for
$\alpha_i = 1 - \beta_i$ with monotonically increasing noise schedule $\beta_1, \beta_2, \dots , \beta_N \in (0, 1)$.
The associated reverse diffusion step is
\begin{align}\label{eq:DDPMr}
\x_{i-1}=\frac{1}{\sqrt{\alpha_i}}\Big(\x_i + {(1 - \alpha_i)}\s_\theta(\x_i, i)\Big) + {\sqrt{\sigma_i}} \z,
\end{align}
where $\s_\theta(\x_i, i)$ is a discrete score function that
matches $\nabla_{\x_i} \log p_{0i}(\x_i|\x_0)$. {Further, the noise term $\sigma_i$ can be fixed to $\sigma_i = 1 - \alpha_i$~\cite{ho2020denoising}, or set to a learnable parameter~\cite{nichol2021improved, dhariwal2021diffusion}.}

For DDPM, denoising diffusion implicit model (DDIM) establishes the current state-of-the-art among the acceleration methods. Unlike DDPM, DDIM has no additive noise term during the reverse diffusion, allowing less iterations for competitive sample quality. Specifically, the reverse diffusion step is given as:
\begin{align}\label{eq:DDIM}
\x_{i-1}&=\sqrt{\bar\alpha_{i-1}}\left(\frac{\x_i-\sqrt{1-\bar\alpha_i}\z_\theta(\x_i,i)}{\sqrt{\bar\alpha_i}}\right)\notag\\
&+\sqrt{1-\bar\alpha_{i-1}}\z_\theta(\x_i,i)
\end{align}
where
\begin{align}
\z_\theta(\x,i) : = -\s_\theta(\x,i){\sqrt{1-\bar\alpha_i}}.
\end{align}
{One can further define}
\begin{equation}
\label{eq:DDIM2}
   {\sigma_i = \frac{\sqrt{1 - \bar\alpha_i}}{\sqrt{\bar\alpha_i}}, \quad \bar\x_i = \frac{\x_i}{\sqrt{\bar\alpha_i}},}
\end{equation}
{to express \eqref{eq:DDIM} as $\bar\x_{i-1} = \bar\x_i + (\sigma_{i-1} - \sigma_i)\z_\theta(\xb_i,i)$.}

 On the other hand, score matching with Langevin dynamic (SMLD) \cite{song2019generative, song2020improved} can be seen
 as the discrete version of VE-SDE.
 Specifically, the forward SMLD diffusion step is given by
 \begin{align}\label{eq:SMLDf}
\x_{i} = {\x}_0 + \sigma_{i}\z
\end{align}
where  $\sigma_i = \sigma_{\text{min}}(\frac{\sigma_{\text{max}}}{\sigma_\text{min}})^{\frac{i-1}{N-1}}$, as defined in~\cite{song2020score}.
The associated reverse diffusion is given by
\begin{align}\label{eq:SMLDr}
\x_{i-1} = \x_i + (\sigma_{i}^2 - \sigma_{i-1}^2)\s_\theta(\x_{i}, i)+  \sqrt{\sigma_{i}^2 - \sigma_{i-1}^2}\z
\end{align}
where $\z \sim \Nc(\textbf{0}, \I)$. 


\section{Main Contribution}
\label{sec:methods}

\subsection{The CCDF Algorithm}

The goal of our CCDF acceleration scheme is to make the reverse diffusion start from $N':=Nt_0<N$ such that
the resulting number of reverse diffusion step can be significantly reduced.
For this, our CCDF algorithm  is composed of two steps: forward diffusion up to $N'$ with {\em better} initialization {$\x_0$},
which is followed by a reverse conditional diffusion down to $i=0$.

Specifically, for a given initial estimate $\x_0$, the forward diffusion process  can be performed {with} a single step diffusion as follows:
\begin{align}
\label{eq:forward_sde}
\x_{N'} = a_{N'}\x_{0}+b_{N'}\z
\end{align}
where $\z \sim \Nc(\textbf{0}, \I)$, and $a_{N'}${,} $b_{N'}$ for SMLD and DDPM can be computed
for each diffusion model using \eqref{eq:SMLDf} and \eqref{eq:DDPMf}, respectively.

In regard to the conditional difusion,
 SRDiff~\cite{li2021srdiff}, SR3~\cite{saharia2021image} are examples that are trained specifically for SR, with the low-resolution counterparts being encoded or concatenated as the input. 
However, these approaches attempt to redesign the score function so that one can sample from the conditional distribution, 
leading to a much complicated formulation.

Instead, here we propose a much simpler but {effective} conditional diffusion.
Specifically, 
our reverse diffusion uses standard reverse diffusion, alternated with an operation to impose data consistency:
\begin{align}
\x'_{i-1} &= \f(\x_i,i)+g(\x_i,i)\z_{i} \label{eq:ourr}\\
\x_{i-1} & = \A\x'_{i-1}+\bb  \label{eq:ourcond}
\end{align}
where the specific forms of $\f(\x_i,i)$ and $g(\x_i,i)$ depend on the {type of} diffusion models (see Table~\ref{tab:notations}),
  $\z_i\sim \Nc(\0,\I)$, and 
$\A$ is a non-expansive mapping \cite{bauschke2011convex}:
\begin{align}\label{eq:nonexp}
\|\A\x-\A\x'\|\leq  \|\x-\x'\|,\quad \forall \x,\x'
\end{align}
 In particular, we assume $\A$ is linear.
For example, one-iteration of the standard gradient descent~\cite{jalal2021robust, ramzi2020denoising}
or projection onto convex sets (POCS) in \cite{tang2011projection, fan2017projections, hosseini2010image, samsonov2004pocsense}
corresponds to our data consistency step in \eqref{eq:ourcond} with \eqref{eq:nonexp}. See Supplementary Section~\ref{sec:supp_algorithm} for algorithms used for each task.

\begin{table}[!hbt]
    \resizebox{0.45\textwidth}{!}{\begin{tabular}{c|c|c}
    & $\fb(\xb_i,i)$ & $g(\xb_i,i)$ \\ \hline\hline
    SMLD & $\xb_i + (\sigma_i^2 - \sigma_{i-1}^2)s_\theta(\xb_i,i)$ & $\sqrt{\sigma_i^2 - \sigma_{i-1}^2}$\\ 
    DDPM & $\frac{1}{\sqrt{\alpha_i}}(\xb_i+(1-\alpha_i)s_\theta(\xb_i,i))$ & $\sqrt{1 - \alpha_i}$\\ 
    DDIM & $\sqrt{\alpha_{i-1}}\left(\frac{\xb_i - \sqrt{1-\bar\alpha_i}z_\theta(\xb_i,i)}{\sqrt{\bar\alpha_i}}\right)+\sqrt{1-\bar\alpha_{i-1}}z_\theta(\xb_i,i)$ &  0\\ \hline
    \end{tabular}}
    \caption{{Values of $\fb, g,$ and noise schedule of discrete SDEs.}}
    \hspace*{-0.5cm}
    \vspace*{-0.5cm}
    \label{tab:notations}
\end{table}

\subsection{Fast Convergence Principle of CCDF}


Now, we are ready to show why CCDF provides much faster convergence than the standard
conditional diffusion models that starts from Gaussian noise.
In fact, the key innovation comes from the mathematical findings that while the
forward diffusion increases the estimation error,
 the conditional reverse diffusion decreases it much faster at exponential rate.
Accordingly,  we can find a ``sweet spot''  $N'$ such that the forward diffusion up to $N'$ followed by reverse
diffusion can significantly reduces the estimation error of the initial estimate $\xb_0$.
This fast convergence principle is shown in the following theorems, whose proofs can be found in Supplementary Materials.
First, the following lemma is a simple consequence of independency of Gaussian noises.

\begin{lemma}
Let $\tilde\x_0\in \Rd^n$ and $\x_0\in\Rd^n$ be the ground-truth clean image and its initial estimate, respectively, and the
initial estimation error is denoted by $\varepsilon_0 = \|\x_0-\tilde\x_0\|^2$. Suppose, furthermore, that
$\x_{N'}$ and $\tilde\x_{N'}$ denote the forward diffusion from $\x_0$ and $\tilde\x_0$, respectively, using  \eqref{eq:forward_sde}.
Then, the estimation error after the forward diffusion is given by
\begin{align}
   \bar\varepsilon_{N'} &:=  \Ed\|\x_{N'} - \tilde\x_{N'}\|^2 \notag\\
   &= a^2_{N'}\varepsilon_0 + 2 b^2_{N'}n.
\end{align}
\end{lemma}

Now, the following theorem, which is a key step of our proof,
 comes from the stochastic contraction property 
of {the} stochastic difference equation \cite{pham2009contraction,pham2008analysis}.
\begin{theorem}
\label{thm:contraction}
Consider the reverse diffusion using \eqref{eq:ourr} and \eqref{eq:ourcond}.
Then, we have
\begin{align}
\bar\varepsilon_{0,r} \leq \frac{2{C}\tau}{1-\lambda^2}+{\lambda}^{2N'} \bar\varepsilon_{N'}
\end{align}
where $\bar\varepsilon_{0,r}$ denotes the {estimation} error between reverse conditional
diffusion path down to $i=0$, and
$\tau = \frac{\mathrm{Tr}(\A^T\A)}{n}$. Furthermore,
the contraction rate $\lambda$ and the {constant} $C$ have the following closed form expression:
\begin{align}
\lambda = \begin{cases}\max\limits_{i\in[N']}\sqrt{\alpha_i}\left(\frac{1-\bar\alpha_{i-1}}{1-\bar\alpha_i}\right)  &\text{(DDPM)}\\ 
\max\limits_{i\in[N']}\frac{\sigma_{i-1}^2 - \sigma_{0}^2}{\sigma_{i}^2 - \sigma_{0}^2} &\text{(SMLD)} \\ 
\max\limits_{i\in[N']}\frac{\sigma_{i-1}}{\sigma_{i}} &\text{(DDIM)} \end{cases}
\end{align}
and
\begin{align}
C= \begin{cases}n(1-\alpha_N)  &\text{(DDPM)}\\ 
{n\max\limits_{i\in[N']}\sigma_i^2-\sigma_{i-1}^2} &\text{(SMLD)} \\ 
0 &\text{(DDIM)} \end{cases}
\end{align}
\end{theorem}
Now we have the main results that shows the existence of the shortcut path for the acceleration.
\begin{theorem}[Shortcut path] 
\label{thm:sp}
For any $0<\mu \leq 1$, there exists a minimum $N' (=t_0N<N)$ such that $\bar\varepsilon_{0,r}\leq\mu\varepsilon_0$.
Furthermore, $N'$ decreases as $\varepsilon_0$ gets smaller.
\end{theorem}
{Theorem~\ref{thm:contraction} states that the conditional reverse diffusion is exponentially contracting. Subsequently, Theorem}~\ref{thm:sp} tells us that we can achieve superior results (i.e. tighter bound) with {\em shorter} sampling path. Hence, it is unnecessary for us to start sampling from $N$. Rather, we can start from an arbitrary timestep $N'<N$,  and still converge faster to the same point that could be achieved when starting the sampling procedure at $N$. 
Furthermore, as we have better initialization such that $\varepsilon_0$ is smaller,
then we need smaller reverse diffusion step, achieving much higher acceleration.

For example,  we can initialize the corrupted image with a pre-trained neural network $G_\varphi$,
which has been  widely studied across different tasks~\cite{ledig2017photo, yu2018generative, yu2019free}. These methods are typically extremely fast to compute, and thus does not introduce additional computational overload. Using this rather simple and fast fix, we observe that we are able to choose smaller values of $t_0$, endowed with much stabler performance. For example, in the case of MRI reconstruction, we can choose $t_0$ as small as 0.02, while {\em outperforming} score-{MRI}~\cite{chung2021score} with 50$\times$ acceleration. 

\section{Experiments}

\subsection{Experimental settings}

We test our method on three different tasks: super-resolution, inpainting, and MRI reconstruction. 
{For all methods, we evaluate the qualitative image quality and quantitative metrics as we accelerate the diffusion process by reducing the $t_0$ values.} For the proposed method, we report on the results starting with neural network (NN)-initialized {$\x_0$} unless specified otherwise.

\noindent
\textbf{Dataset.}~For vision tasks using face images, we use two datasets - FFHQ $256 \times 256$, and AFHQ $256 \times 256$. For FFHQ, we randomly select 50k images for training, and sample 1k images of test data separately. For AFHQ, we train our model using the images in the \code{dog} category, which consists of about 5k images. Testing was performed with the held-out validation set of 500 images of the same category. For the MRI reconstruction task, we use the fastMRI knee data, which consists of around 30k $320\times320$-sized slices of coronal knee scans. Specifically, we use magnitude data given as the key \code{reconstruction\_esc}. We randomly sample 10 volumes from the validation set for testing.

\noindent
\textbf{Quantitative metrics.}~Since it is well known that for high corruption factors, standard metrics such as PSNR/SSIM does not correlate well with the visual quality of the reconstruction~\cite{saharia2021image, yu2019free}, we report on the FID score based on \code{pytorch-fid}\footnote{\href{https://github.com/mseitzer/pytorch-fid}{https://github.com/mseitzer/pytorch-fid}}. For MRI reconstruction, it is less sound to report on FID; hence, we report on PSNR.

\begin{figure}[!t]
    \centering\includegraphics[width=8cm]{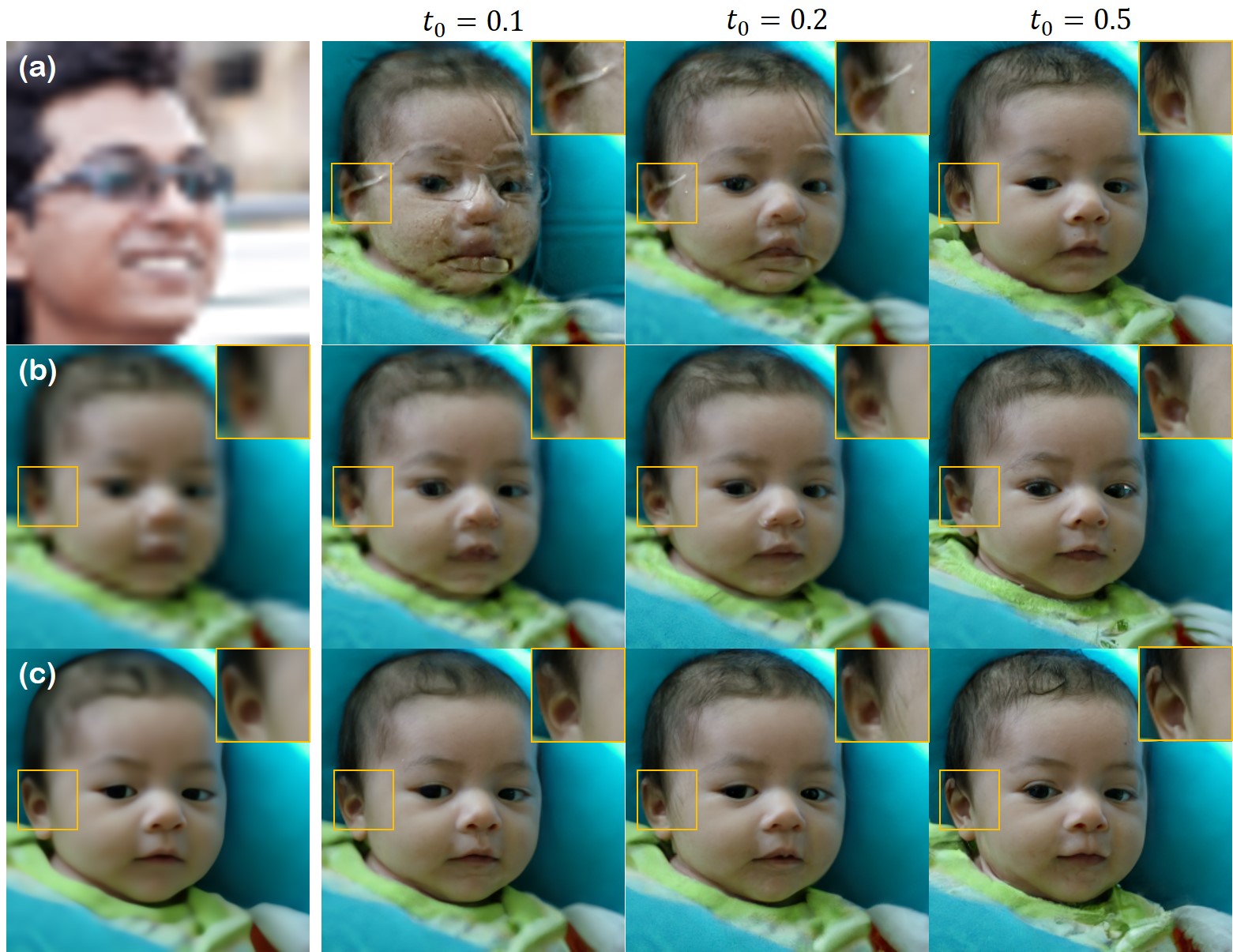}
    \caption{{Stability of convergence depending on the choice of initialization. (a) Random initialization, large $\varepsilon_0$, (b) vanilla initialization, moderate $\varepsilon_0$, (c) NN initialization, small $\varepsilon_0$.}}
	\label{fig:toy_example}
\end{figure}

\begin{figure*}[!hbt]
    \centering\includegraphics[width=17cm]{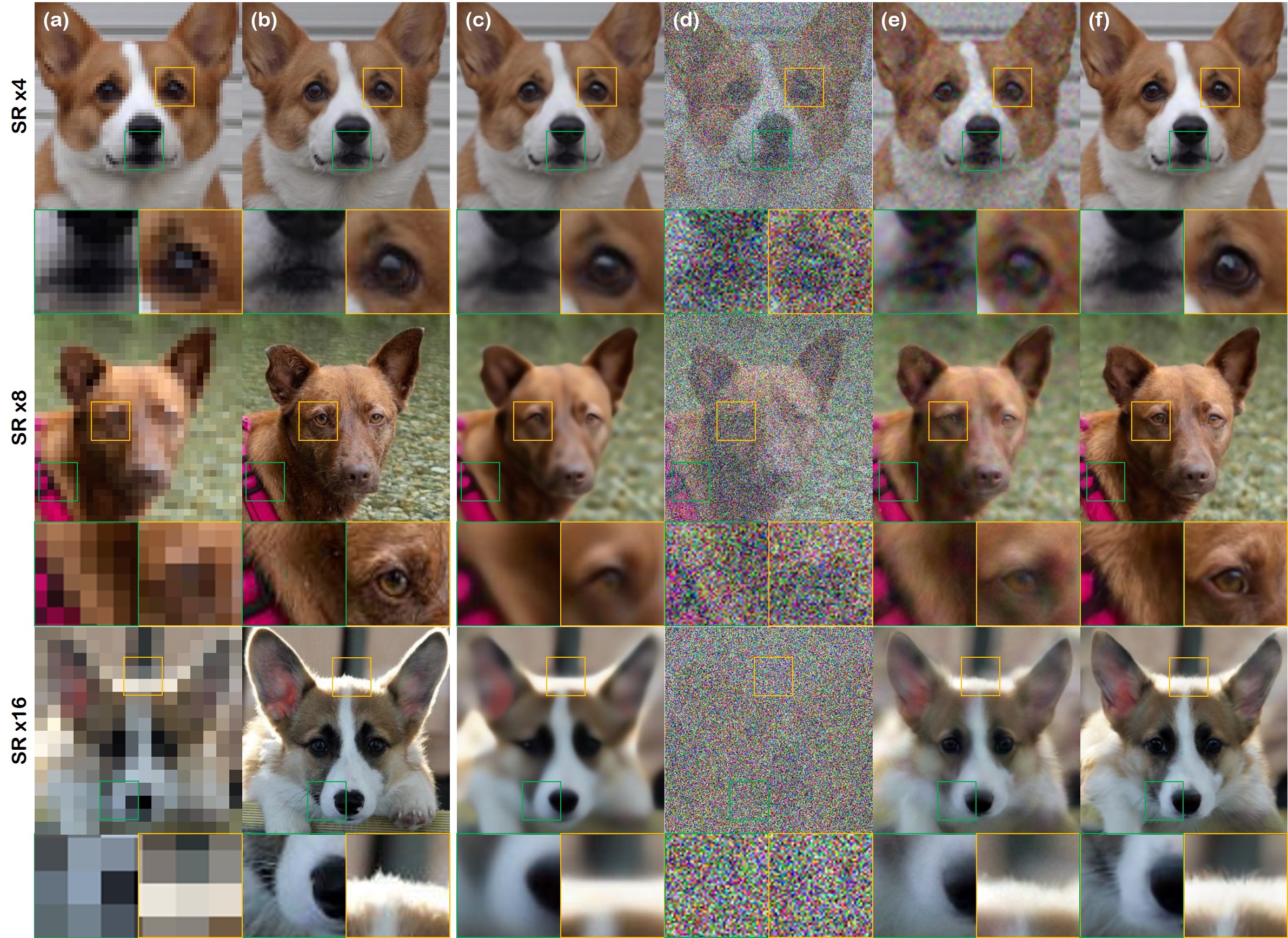}
    \caption{Results of super-resolution on AFHQ 256$\times$256 data. First, second and third row denote $\times 4$ SR, $\times 8$ SR, and $\times 16$ SR, respectively. (a) LR input, (b) Ground Truth, (c) ESRGAN~\cite{wang2018esrgan}, (d) SR3~\cite{saharia2021image} with 20 diffusion steps {($N = 20, \Delta t = 0.05$)}, (e) ILVR~\cite{choi2021ilvr} with 20 diffusion steps {($N = 20, \Delta t = 0.05$)}, (f) proposed method with 20 diffusion steps ($N = 100, t_0 = 0.2$).}
	\label{fig:SR_results}
\end{figure*}

\noindent
\textbf{Super-resolution.}~Experiments were performed across three different levels of SR factor - $\times 4, \times 8, \times 16$.
We train a discretized VP-SDE based on IDDPM~\cite{nichol2021improved} for each dataset - FFHQ and AFHQ, following the standards. Specific details can be found in Supplementary section \ref{sec:supp_algorithm}. For the one-step feed forward network corrector, we train the widely-used ESRGAN~\cite{wang2018esrgan} for each SR factor, using the same neural network architecture that was used to train the score function. We use three methods for comparison - ESRGAN, ILVR, and SR3\footnote{\href{https://github.com/Janspiry/Image-Super-Resolution-via-Iterative-Refinement}{https://github.com/Janspiry/Image-Super-Resolution-via-Iterative-Refinement}}. We note that the official code of SR3 is yet to be released, and hence we resort to unofficial re-implementation, which we train with default configurations. {Additionally, in the original work of SR3~\cite{saharia2021image}, the authors propose consecutively applying $\times4$ SR models to achieve $16\times16 \mapsto 64\times64 \mapsto 256\times256$ SR. In contrast, we report on a single $\times 16$ SR model which maps $16\times16 \mapsto 256\times256$ directly.}

\noindent
\textbf{Inpainting.}~The score function used in the inpainting task is the same model that was used to solve SR tasks, since we use task-agnostic conditional diffusion model. The feed-forward network was adopted from Yu {\em et al.}~\cite{yu2019free}. We consider box-type inpainting with varying sizes: $96\times96, 128\times128, 160\times160$. The model was trained for 50k steps with default configurations. We compare with score-SDE~\cite{song2020score}, using the same trained score function.

\noindent
\textbf{MRI reconstruction.}~Experiments were performed across three different levels of acceleration factor, with gaussian 1D sampling pattern - $\times 2, \times 4, \times 6$, each with 10\%, 8\%, 6\% of the phase encoding lines included for autocalibrating signal (ACS) region.
We train a VE-SDE based on \code{ncsnpp}, proposed in~\cite{song2020score}, and demonstrated specifically for MR reconstruction in~\cite{chung2021score, song2022solving}. For comparison with compressed sensing (CS) strategy, we use total-variation (TV) regularized reconstruction. For feed forward network, we train a standard U-Net, using similar settings from \cite{chung2021score, zbontar2018fastmri}. We use the same trained score function for comparison with score-MRI~\cite{chung2021score}.

\subsection{Super-resolution}
\label{sec:SR}

\noindent
\textbf{{Dependence on $\varepsilon_0$.}}~We first demonstrate the dependency of stochastic contraction on the squared error term in Figure~\ref{fig:toy_example}. For small squared difference, as in the case for many inverse problems, we see that the reverse diffusion stably converges to the same solution, even with small timestep $t_0$. In contrast, when random {$\x_0$} is the starting point, $\varepsilon_0$ becomes large, and only with higher values of $t_0$ does the reverse SDE converge to a feasible solution.

\begin{table}[!hbt]
    \centering
    \resizebox{0.45\textwidth}{!}{
    \begin{tabular}{c|cccccc}
    \hline
         $t_0$ & 0.05 & 0.1 & 0.2 & 0.5 & 0.75 & 1.0~\cite{choi2021ilvr} \\ \hline\hline
    SR $\times 4$ & 63.90 & \textbf{60.90} & \underline{60.91} & 64.04 & 64.14 & 63.31 \\ 
    SR $\times 8$ & 85.21 & 78.13 & \textbf{75.76} & 79.34 & 79.67 & \underline{77.34} \\ 
    SR $\times 16$ & 116.37 & 101.79 & 92.59 & \textbf{88.09} & 92.12 & \underline{88.49} \\ \hline
    \end{tabular}
    }
    \caption{FID($\downarrow$) scores on FFHQ test set for SR task with $N = 1000$, and varying $t_0$ values. $t_0 = 1.0$ is the baseline method without any acceleration used in~\cite{choi2021ilvr}. Numbers in boldface and underline indicate the best and the second best.}
    \label{tab:FID_SR_FFHQ}
\end{table}

\noindent
\textbf{{Dependence on $t_0$.}}~In Table~\ref{tab:FID_SR_FFHQ}, we report on the FID scores by varying the $t_0$ values with a fixed discretization step $\Delta t = 1/1000$ in order  to see which value is optimal for each degradation factor. Consistent with the theoretical findings, we see that as the corruption factor gets higher, and $\varepsilon_0$ gets larger, we typically need higher values of $t_0$ to achieve optimal results. Interesting enough, we observe that there {\em always} exist a value $t_0 \in [0, 1)$ where the FID score is lower (lower is better) than when using full reverse diffusion from $T=1$.

\begin{figure}[!hbt]
    \centering\includegraphics[width=6cm]{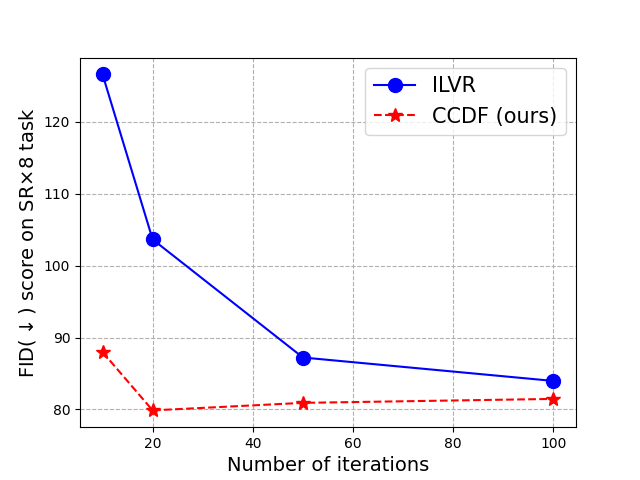}
    \caption{Comparison of FID score on $\times$8 SR task. For ILVR, re-scheduling method of IDDPM~\cite{nichol2021improved} was used starting from $T=1$. For {CCDF}, the step size for discretization is $\Delta t=0.01$ so that the starting point for the
    reverse diffusion is  $t_0 = \Delta t \times \mbox{[number of iteration]}$. }
	\label{fig:plot_comparison}
\end{figure}

\noindent
\textbf{{Comparison study.}}~The results of various super-resolution algorithms is compared in Fig.~\ref{fig:SR_results}. We compare with {SR3~\cite{saharia2021image} and} ILVR~\cite{choi2021ilvr}, with setting the number of iterations for reconstruction same for ILVR, {SR3}, and the proposed method. We clearly see that {SR3 and} ILVR starting from pure Gaussian noise at $T=1$  cannot generate satisfactory results with 20 iterations, whereas our method can estimate high-fidelity samples with details preserved {even} with only 20 iterations starting from $t_0=0.2$. Visualizing the trend of FID score in Figure~\ref{fig:plot_comparison}, we see that the quality of the image degrades as we use less and less number of iterations for the ILVR method, whereas the proposed method is able to keep the FID score at about the same level, or even {\em boost} the image quality, with less iterations.
%
\vspace{-0.5cm}
\begin{table}[!hbt]
    \centering
    \resizebox{0.45\textwidth}{!}{
    \begin{tabular}{c|c|cccc}
    \hline
    & SR factor & ESRGAN\cite{wang2018esrgan} & SR3$^{*}$\cite{saharia2021image} & ILVR\cite{choi2021ilvr} & {CCDF (ours)} \\ \hline\hline
        \multirow{3}{*}{FFHQ} & $\times$4 &81.14&66.79&63.14&\textbf{60.90}  \\ 
        & $\times$8 &108.96&80.27&81.85&\textbf{75.76} \\ 
        & $\times$16 &143.80&99.46&92.32&\textbf{88.39} \\ \hline
        \multirow{3}{*}{AFHQ} & $\times$4 & 24.52&20.68&18.70&\textbf{15.53} \\
        & $\times$8 &51.84&30.23&34.85&\textbf{32.30} \\
        & $\times$16 &98.22&60.76&\textbf{47.28}&48.77 \\ \hline
    \end{tabular}
    }
    \caption{Comparison of FID($\downarrow$) scores on FFHQ and AFHQ test set. $t_0$ values used for the proposed method is $0.1, 0.2, 0.3$ for $\times4, \times8, \times16$ SR, respectively. Numbers in boldface represent the best results among the row. ($^{*}$unofficial re-implementation)}
    \label{tab:SR_comparison}
\end{table}

\begin{figure}[!hbt]
    \centering\includegraphics[width=8cm]{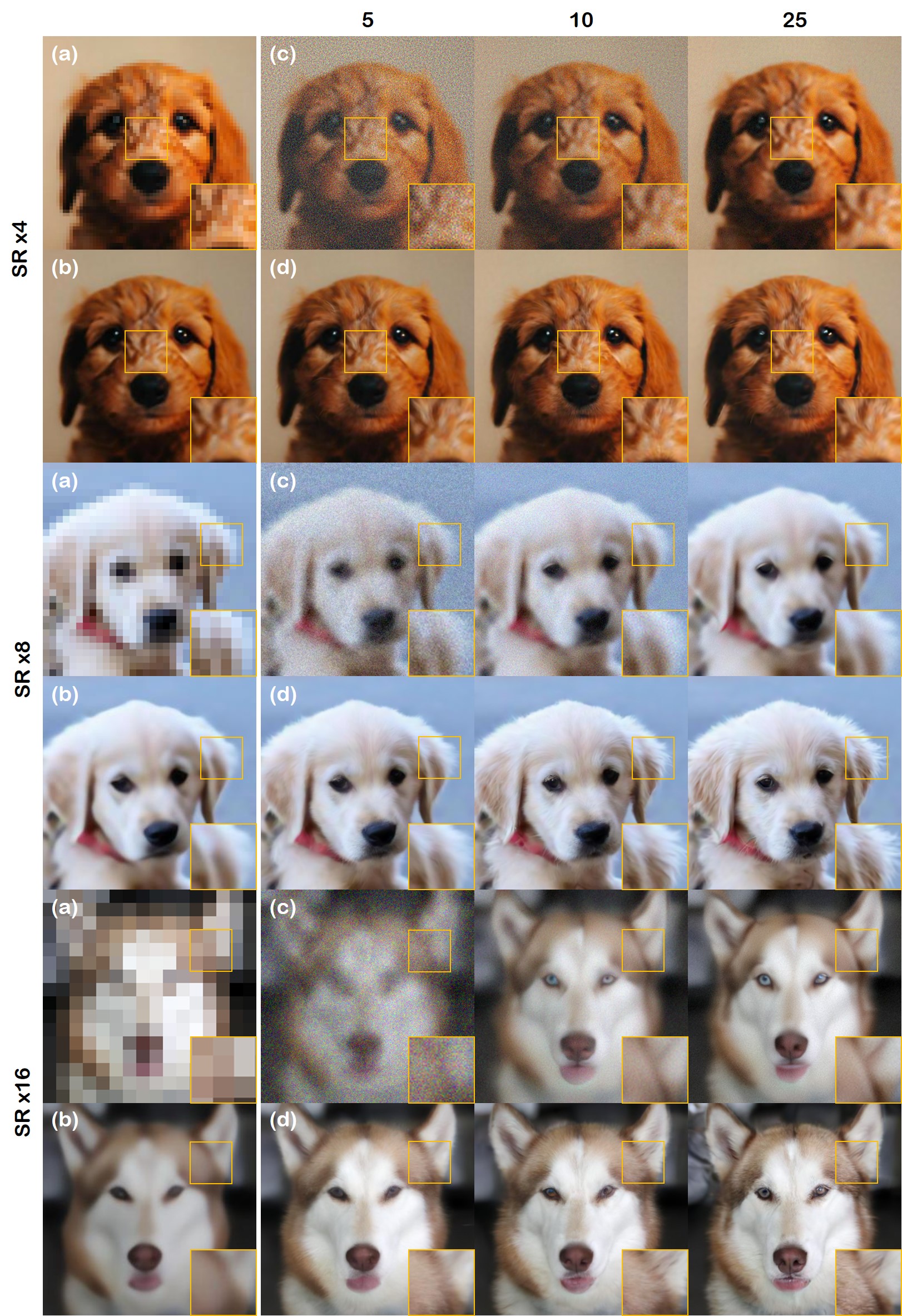}
    \caption{{Results on SR task using CCDF with DDIM. (a) LR image, (b) initialization with ESRGAN, (c) ILVR + DDIM, (d) CCDF + DDIM. Numbers on top indicate the number of iterations. Proposed method uses $N = 50$, and $t_0 = 0.1, 0.2, 0.5$, respectively.}}
	\label{fig:ddim_sr_main}
\end{figure}

We also perform a comparison study where we set the total number of diffusion steps to $N = 1000$ starting from $T=1$ for ILVR~\cite{choi2021ilvr},
 and set $t_0$ to 0.1, 0.2, and 0.3 for each factor, thereby reducing the
 number of diffusion steps to 100, 200, and 300, respectively, by our method. In Table~\ref{tab:SR_comparison}, we demonstrate by using the proposed method, we achieve results that are on par or even better. {For qualitative analysis, see Supplementary Section~\ref{sec:supp_ae}.}
 
\noindent
\textbf{{Incorporation of DDIM.}}~{As briefly discussed before, CCDF can be combined together with approaches that searches for the optimal (full) reverse diffusion path. In Fig.~\ref{fig:ddim_sr_main}, we illustrate that we can reduce the number of iterations to as little as {\em 5 steps}, and still maintain high image quality.}

\subsection{Inpainting}
\label{sec:inpaint}


\begin{figure}[!hbt]
    \centering\includegraphics[width=8cm]{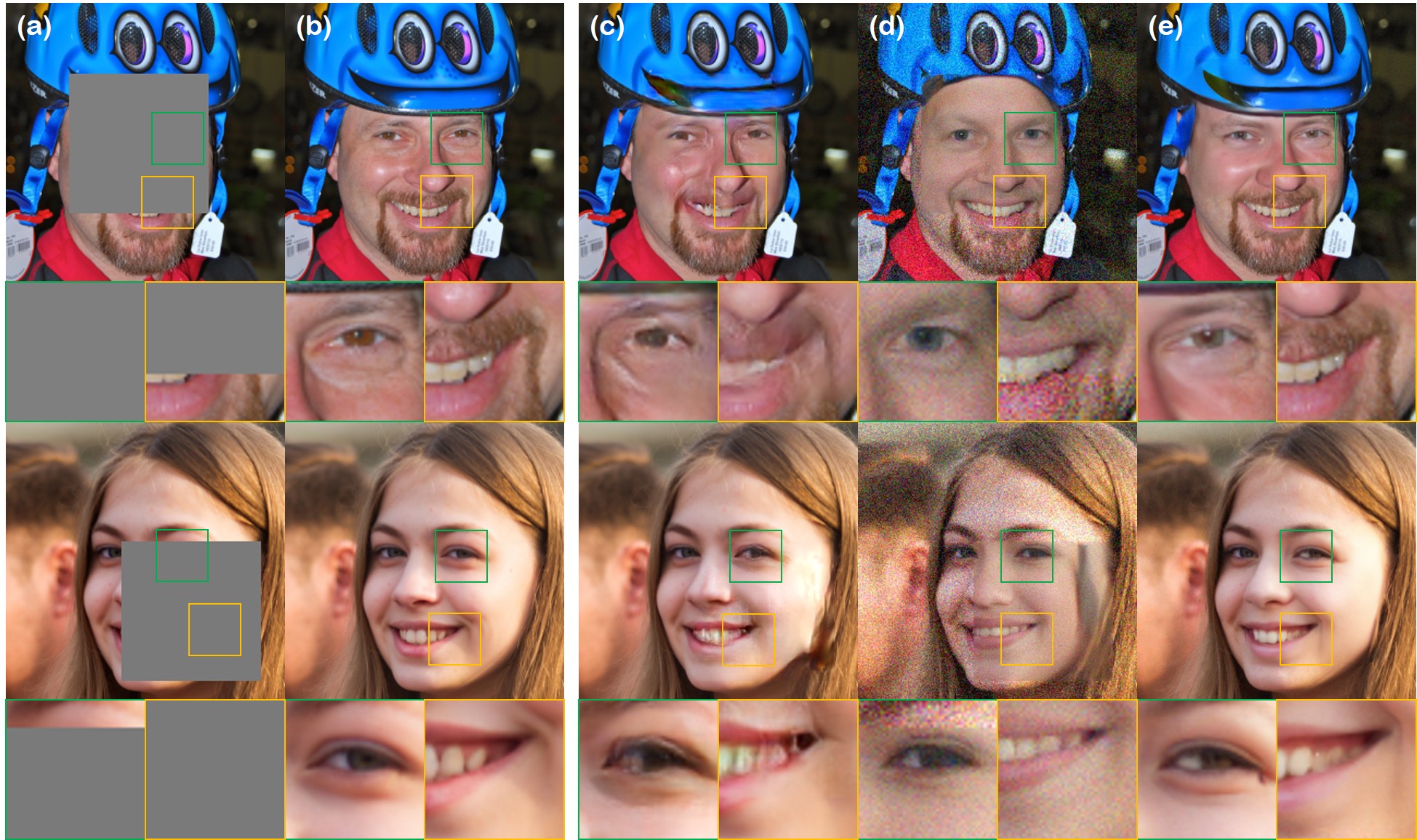}
    \caption{Results of inpainting on FFHQ 256$\times$256 data. (a) Masked image, (b) Ground Truth, (c) SN-PatchGAN~\cite{yu2019free}. (d) score-SDE~\cite{song2020score} using 20 steps from $T=1$, (e) proposed method {(CCDF)} using $20$ steps from $t_0 = 0.2$.}
	\label{fig:inpaint_results_ffhq1}
\end{figure}
We illustrate the results of inpainting in Fig.~\ref{fig:inpaint_results_ffhq1}. 
Consistent with what was observed in the SR task, the results in Figure~\ref{fig:inpaint_results_ffhq1} show that using full reverse diffusion with large discretization steps is inefficient, leading to unrealistic output. On the other hand, our method can reconstruct very realistic images within this small budget.

\begin{table}[!hbt]
    \centering
    \resizebox{0.45\textwidth}{!}{
    \begin{tabular}{c|cccc}
    \hline
    method & masked & \thead{SN-PatchGAN\\\cite{yu2019free}} & \thead{Score-SDE\cite{song2020score}\\(1000)} & \thead{{CCDF}\\(200)} \\ \hline\hline
    Box 96 & 131.31 & 46.42 & 50.85 & \textbf{45.99} \\ 
    Box 128 & 145.81 & 52.63 & 64.51 & \textbf{49.77} \\
    Box 160 & 167.37 & 66.25 & 78.29 & \textbf{57.99} \\ \hline
    \end{tabular}
    }
    \caption{Comparison of FID($\downarrow$) scores on FFHQ test set for inpainting task {($N=1000, \Delta t = 0.001$)}. Number in parenthesis indicate the number of iterations used for generation.
    Numbers in boldface and underline indicate the best and the second best.}
    \label{tab:FID_inpaint_FFHQ_comparison}
    \vspace{-0.5cm}
\end{table}

Comparison with prior arts by setting relatively large number of iterations is shown in Table~\ref{tab:FID_inpaint_FFHQ_comparison}. {We observe that the proposed method outperforms both score-SDE with full reverse diffusion, and SN-PatchGAN, in terms of FID score. For detailed comparison and further experiments, see Supplementary Section~\ref{sec:supp_ae}.}

\subsection{MRI reconstruction}

\begin{figure}[!hbt]
    \centering\includegraphics[width=8cm]{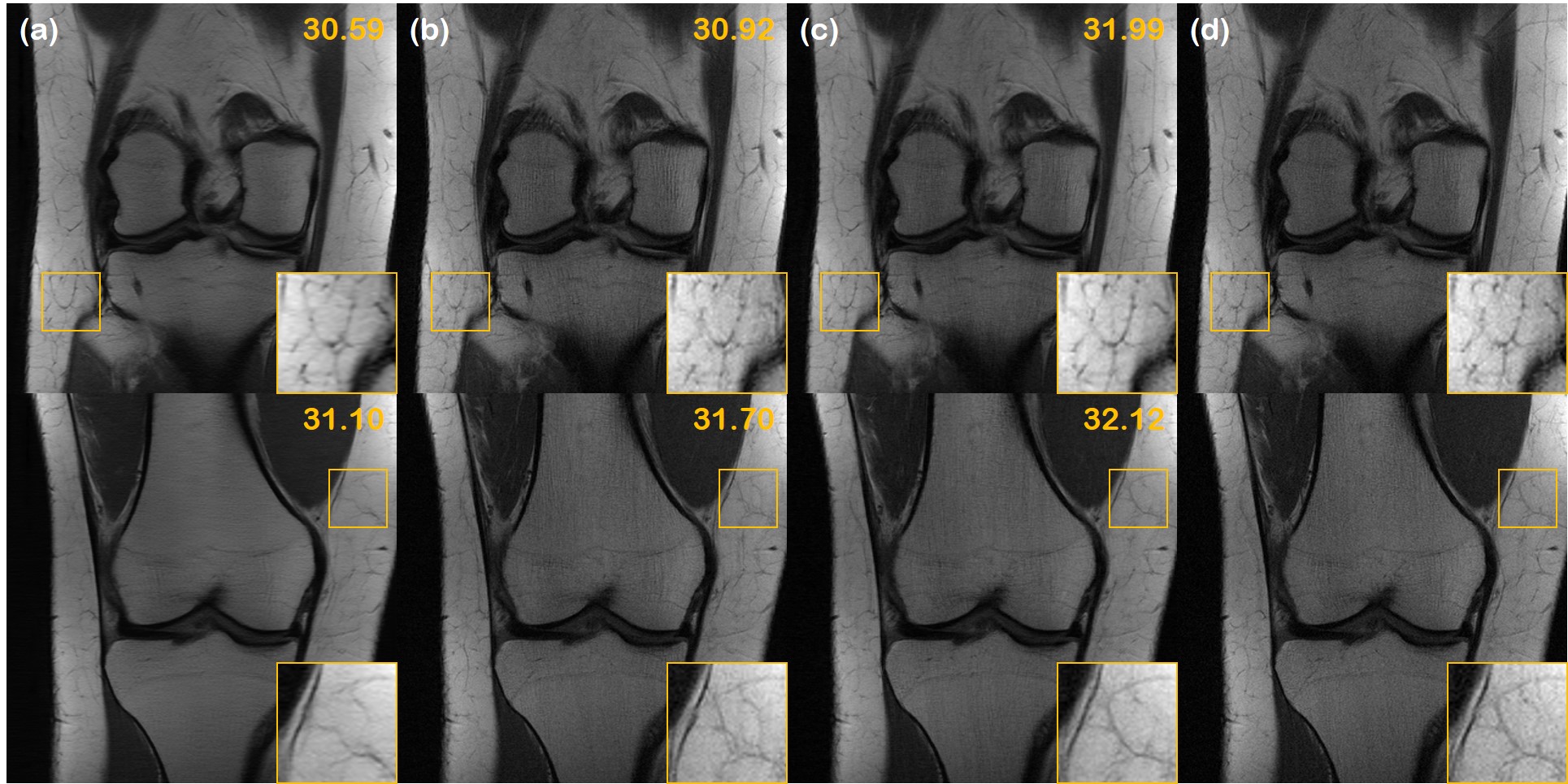}
    \caption{Results of the MR reconstruction task: (a) TV~\cite{block2007undersampled}, (b) U-Net~\cite{zbontar2018fastmri}, (c) score-POCS~\cite{chung2021score} using 1000 steps starting from  $T=1$, (d) proposed method {(CCDF)} using 20 steps from $t_0 = 0.02$ (20 steps), (e) Reference image. Numbers in yellow correspond to PSNR values.}
	\label{fig:MR_comparison}
	\vspace{-0.6cm}
\end{figure}

We summarize and compare our results in Figure~\ref{fig:MR_comparison}, and the quantitative metrics are presented in Table~\ref{tab:MR_recon}. In the task of MR reconstruction, we observe that we can push the $t_0$ value down to very small values: $t_0 = 0.02$, and still achieve remarkable results, even outperforming score-POCS which uses full reverse diffusion. When we compare the proposed method which uses 20 iterations vs. score-POCS with 20 iterations, we see that score-POCS cannot generate a feasible image, arriving at what looks like pure noise, as demonstrated in Figure~\ref{fig:concept}.
With other tasks, we could see that higher degradations typically require increased $t_0$ values. With CCDF, we do not see such trend, and observe that selecting low values of $t_0 \in [0.02, 0.1]$ stably gives good results.
We emphasize that this is a huge leap towards practical usage of diffusion models in clinical settings, where fast reconstruction is crucial for real-time deployment.

\begin{table}[!hbt]
    \centering
    \resizebox{0.5\textwidth}{!}{
    \begin{tabular}{c|ccccc}
    \hline
    method & ZF & TV~\cite{block2007undersampled} & U-Net~\cite{zbontar2018fastmri} & \thead{Score-POCS \\ \cite{chung2021score}} & \thead{{CCDF} \\ ({20})} \\ \hline\hline
    $\times$ 2 & 27.23 & 29.10 & 32.93 & 32.85 & \textbf{33.41} \\ 
    $\times$ 4 & 22.68 & 25.93 & 31.07 & 31.45 & \textbf{32.51} \\
    $\times$ 6 & 21.54 & 24.69 & 30.77 & 31.15 & \textbf{31.30} \\ \hline
    \end{tabular}
    }
    \caption{PSNR($\uparrow$) on fastMRI test set for MRI reconstruction tasks. Gaussian 1D sampling masks were used. {Number in parenthesis indicate the number of iterations used ($N=1000, t_0 = 0.02$}). Numbers in boldface indicate the best  among the rows.}
    \label{tab:MR_recon}
    \vspace{-0.3cm}
\end{table}

\section{Discussion}
We note that we are not the first to propose starting from forward-diffused data in the context of diffusion models. It was first introduced in SDEdit~\cite{meng2021sdedit}, but in a different context with distinct aim form ours. In SDEdit, forward diffusion was used up to $t_0 \in [0.3, 0.6]$, which a relatively higher value than those used in our work $t_0 \leq 0.2$, since the purpose was to {\em destroy} the signal so as to acquire high fidelity images from coarse strokes.

Our work differs from SDEdit in that we consider this procedure in a more rigorous framework and first {reveal} that
starting from a  better initialization for inverse problems significantly {accelerate} the reverse diffusion.
This leads to a novel hybridization that has not been covered before: a simple incorporation of pre-trained feed-forward NNs can be very efficient at pushing $t_0$ to smaller limits, even as small as $t_0 = 0.02$ in the case of MRI reconstruction.
%
%

\subsection{Limitations}
We note that the choice of $t_0$ for acceleration varies by quite a margin across different tasks, and the degree of corruptions. Currently, there does not exist clear and concise rules for selecting such values as we do not have a knowledge of $\varepsilon_0$  a priori.
Thus, one needs to rely mostly on trial-and-error, which could potentially reduce practicality. Building an adaptive method that can automatically search for the optimal $t_0$ values will be beneficial, and we leave this venue for possible direction of future research.

\section{Conclusion}

In this work, we proposed a method to accelerate conditional diffusion models, by studying the property of stochastic contraction. When solving inverse problems via conditional reverse diffusion, rather than starting at random Gaussian noise, we proposed to initialize the starting from forward-diffused data from a better initialization, such as one-step correction via NN. 
Using the stochastic contraction theory, we showed theoretically why taking the shortcut path is in fact optimal, and back our statement by showing diverse applications in which we both achieve acceleration along with increased stability and performance.

{\small
\bibliographystyle{ieee_fullname}
\bibliography{egbib}

\begin{thebibliography}{10}\itemsep=-1pt

\bibitem{alain2014regularized}
Guillaume Alain and Yoshua Bengio.
\newblock What regularized auto-encoders learn from the data-generating
  distribution.
\newblock {\em The Journal of Machine Learning Research}, 15(1):3563--3593,
  2014.

\bibitem{bauschke2011convex}
Heinz~H Bauschke, Patrick~L Combettes, et~al.
\newblock {\em Convex analysis and monotone operator theory in {H}ilbert
  spaces}, volume 408.
\newblock Springer, 2011.

\bibitem{block2007undersampled}
Kai~Tobias Block, Martin Uecker, and Jens Frahm.
\newblock {Undersampled radial MRI with multiple coils. Iterative image
  reconstruction using a total variation constraint}.
\newblock {\em Magnetic Resonance in Medicine: An Official Journal of the
  International Society for Magnetic Resonance in Medicine}, 57(6):1086--1098,
  2007.

\bibitem{chen2021wavegrad}
Nanxin Chen, Yu Zhang, Heiga Zen, Ron~J Weiss, Mohammad Norouzi, and William
  Chan.
\newblock {WaveGrad}: Estimating gradients for waveform generation.
\newblock In {\em International Conference on Learning Representations}, 2021.

\bibitem{choi2021ilvr}
Jooyoung Choi, Sungwon Kim, Yonghyun Jeong, Youngjune Gwon, and Sungroh Yoon.
\newblock {ILVR}: Conditioning method for denoising diffusion probabilistic
  models.
\newblock In {\em Proceedings of the IEEE/CVF International Conference on
  Computer Vision (ICCV)}, 2021.

\bibitem{chung2021score}
Hyungjin Chung and Jong~Chul Ye.
\newblock Score-based diffusion models for accelerated mri.
\newblock {\em arXiv preprint arXiv:2110.05243}, 2021.

\bibitem{deng2009imagenet}
Jia Deng, Wei Dong, Richard Socher, Li-Jia Li, Kai Li, and Li Fei-Fei.
\newblock Imagenet: A large-scale hierarchical image database.
\newblock In {\em 2009 IEEE conference on computer vision and pattern
  recognition}, pages 248--255. Ieee, 2009.

\bibitem{dhariwal2021diffusion}
Prafulla Dhariwal and Alex Nichol.
\newblock Diffusion models beat {GAN}s on image synthesis.
\newblock In {\em Advances in Neural Information Processing Systems}, 2021.

\bibitem{fan2017projections}
Chong Fan, Chaoyun Wu, Grand Li, and Jun Ma.
\newblock Projections onto convex sets super-resolution reconstruction based on
  point spread function estimation of low-resolution remote sensing images.
\newblock {\em Sensors}, 17(2):362, 2017.

\bibitem{ho2020denoising}
Jonathan Ho, Ajay Jain, and Pieter Abbeel.
\newblock Denoising diffusion probabilistic models.
\newblock In {\em Advances in Neural Information Processing Systems},
  volume~33, pages 6840--6851, 2020.

\bibitem{hosseini2010image}
Hossein Hosseini, Neda~Barzegar Marvasti, and Farrokh Marvasti.
\newblock Image inpainting using sparsity of the transform domain.
\newblock {\em arXiv preprint arXiv:1011.5458}, 2010.

\bibitem{huang2021variational}
Chin-Wei Huang, Jae~Hyun Lim, and Aaron Courville.
\newblock A variational perspective on diffusion-based generative models and
  score matching.
\newblock In {\em Advances in Neural Information Processing Systems}, 2021.

\bibitem{jalal2021robust}
Ajil Jalal, Marius Arvinte, Giannis Daras, Eric Price, Alexandros~G Dimakis,
  and Jonathan Tamir.
\newblock Robust compressed sensing mri with deep generative priors.
\newblock {\em Advances in Neural Information Processing Systems}, 34, 2021.

\bibitem{kingma2014adam}
Diederik~P. Kingma and Jimmy Ba.
\newblock Adam: {A} method for stochastic optimization.
\newblock In {\em 3rd International Conference on Learning Representations,
  {ICLR}}, 2015.

\bibitem{kingma2021variational}
Diederik~P Kingma, Tim Salimans, Ben Poole, and Jonathan Ho.
\newblock Variational diffusion models.
\newblock {\em Advances in Neural Information Processing Systems}, 34, 2021.

\bibitem{ledig2017photo}
Christian Ledig, Lucas Theis, Ferenc Husz{\'a}r, Jose Caballero, Andrew
  Cunningham, Alejandro Acosta, Andrew Aitken, Alykhan Tejani, Johannes Totz,
  Zehan Wang, et~al.
\newblock Photo-realistic single image super-resolution using a generative
  adversarial network.
\newblock In {\em Proceedings of the IEEE conference on computer vision and
  pattern recognition}, pages 4681--4690, 2017.

\bibitem{li2021srdiff}
Haoying Li, Yifan Yang, Meng Chang, Huajun Feng, Zhihai Xu, Qi Li, and Yueting
  Chen.
\newblock {SRDiff}: Single image super-resolution with diffusion probabilistic
  models.
\newblock {\em arXiv preprint arXiv:2104.14951}, 2021.

\bibitem{luhman2021knowledge}
Eric Luhman and Troy Luhman.
\newblock Knowledge distillation in iterative generative models for improved
  sampling speed.
\newblock {\em arXiv preprint arXiv:2101.02388}, 2021.

\bibitem{meng2021sdedit}
Chenlin Meng, Yang Song, Jiaming Song, Jiajun Wu, Jun-Yan Zhu, and Stefano
  Ermon.
\newblock Sdedit: Image synthesis and editing with stochastic differential
  equations.
\newblock {\em CoRR}, abs/2108.01073, 2021.

\bibitem{nichol2021improved}
Alexander~Quinn Nichol and Prafulla Dhariwal.
\newblock Improved denoising diffusion probabilistic models.
\newblock In {\em Proceedings of the 38th International Conference on Machine
  Learning}, volume 139 of {\em Proceedings of Machine Learning Research},
  pages 8162--8171. PMLR, 2021.

\bibitem{park2021generative}
Sung~Woo Park, Dong~Wook Shu, and Junseok Kwon.
\newblock Generative adversarial networks for markovian temporal dynamics:
  Stochastic continuous data generation.
\newblock In {\em International Conference on Machine Learning}, pages
  8413--8421. PMLR, 2021.

\bibitem{pham2008analysis}
Quang-Cuong Pham.
\newblock Analysis of discrete and hybrid stochastic systems by nonlinear
  contraction theory.
\newblock In {\em 2008 10th International Conference on Control, Automation,
  Robotics and Vision}, pages 1054--1059. IEEE, 2008.

\bibitem{pham2009contraction}
Quang-Cuong Pham, Nicolas Tabareau, and Jean-Jacques Slotine.
\newblock A contraction theory approach to stochastic incremental stability.
\newblock {\em IEEE Transactions on Automatic Control}, 54(4):816--820, 2009.

\bibitem{ramzi2020denoising}
Zaccharie Ramzi, Benjamin Remy, Francois Lanusse, Jean-Luc Starck, and Philippe
  Ciuciu.
\newblock Denoising score-matching for uncertainty quantification in inverse
  problems.
\newblock In {\em NeurIPS 2020 Workshop on Deep Learning and Inverse Problems},
  2020.

\bibitem{saharia2021image}
Chitwan Saharia, Jonathan Ho, William Chan, Tim Salimans, David~J Fleet, and
  Mohammad Norouzi.
\newblock Image super-resolution via iterative refinement.
\newblock {\em arXiv preprint arXiv:2104.07636}, 2021.

\bibitem{salimans2022progressive}
Tim Salimans and Jonathan Ho.
\newblock Progressive distillation for fast sampling of diffusion models.
\newblock In {\em International Conference on Learning Representations}, 2022.

\bibitem{samsonov2004pocsense}
Alexei~A Samsonov, Eugene~G Kholmovski, Dennis~L Parker, and Chris~R Johnson.
\newblock {POCSENSE: POCS-based reconstruction for sensitivity encoded magnetic
  resonance imaging}.
\newblock {\em Magnetic Resonance in Medicine: An Official Journal of the
  International Society for Magnetic Resonance in Medicine}, 52(6):1397--1406,
  2004.

\bibitem{sasaki2021unit}
Hiroshi Sasaki, Chris~G Willcocks, and Toby~P Breckon.
\newblock {UNIT-DDPM}: Unpaired image translation with denoising diffusion
  probabilistic models.
\newblock {\em arXiv preprint arXiv:2104.05358}, 2021.

\bibitem{sohl2015deep}
Jascha Sohl-Dickstein, Eric Weiss, Niru Maheswaranathan, and Surya Ganguli.
\newblock Deep unsupervised learning using nonequilibrium thermodynamics.
\newblock In {\em International Conference on Machine Learning}, pages
  2256--2265. PMLR, 2015.

\bibitem{song2020denoising}
Jiaming Song, Chenlin Meng, and Stefano Ermon.
\newblock Denoising diffusion implicit models.
\newblock In {\em 9th International Conference on Learning Representations,
  {ICLR}}, 2021.

\bibitem{song2019generative}
Yang Song and Stefano Ermon.
\newblock Generative modeling by estimating gradients of the data distribution.
\newblock In {\em Advances in Neural Information Processing Systems},
  volume~32, 2019.

\bibitem{song2020improved}
Yang Song and Stefano Ermon.
\newblock Improved techniques for training score-based generative models.
\newblock In {\em Advances in Neural Information Processing Systems},
  volume~33, pages 12438--12448, 2020.

\bibitem{song2022solving}
Yang Song, Liyue Shen, Lei Xing, and Stefano Ermon.
\newblock Solving inverse problems in medical imaging with score-based
  generative models.
\newblock In {\em International Conference on Learning Representations}, 2022.

\bibitem{song2020score}
Yang Song, Jascha Sohl{-}Dickstein, Diederik~P. Kingma, Abhishek Kumar, Stefano
  Ermon, and Ben Poole.
\newblock Score-based generative modeling through stochastic differential
  equations.
\newblock In {\em 9th International Conference on Learning Representations,
  {ICLR}}, 2021.

\bibitem{tang2011projection}
Zhifei Tang, Mike Deng, Chuangbai Xiao, and Jing Yu.
\newblock Projection onto convex sets super-resolution image reconstruction
  based on wavelet bi-cubic interpolation.
\newblock In {\em Proceedings of 2011 International Conference on Electronic \&
  Mechanical Engineering and Information Technology}, volume~1, pages 351--354.
  IEEE, 2011.

\bibitem{vaswani2017attention}
Ashish Vaswani, Noam Shazeer, Niki Parmar, Jakob Uszkoreit, Llion Jones,
  Aidan~N Gomez, {\L}ukasz Kaiser, and Illia Polosukhin.
\newblock Attention is all you need.
\newblock In {\em Advances in neural information processing systems}, pages
  5998--6008, 2017.

\bibitem{wang2018esrgan}
Xintao Wang, Ke Yu, Shixiang Wu, Jinjin Gu, Yihao Liu, Chao Dong, Yu Qiao, and
  Chen Change~Loy.
\newblock Esrgan: Enhanced super-resolution generative adversarial networks.
\newblock In {\em Proceedings of the European conference on computer vision
  (ECCV) workshops}, pages 0--0, 2018.

\bibitem{watson2021learning}
Daniel Watson, Jonathan Ho, Mohammad Norouzi, and William Chan.
\newblock Learning to efficiently sample from diffusion probabilistic models.
\newblock {\em arXiv preprint arXiv:2106.03802}, 2021.

\bibitem{yu2018generative}
Jiahui Yu, Zhe Lin, Jimei Yang, Xiaohui Shen, Xin Lu, and Thomas~S Huang.
\newblock Generative image inpainting with contextual attention.
\newblock In {\em Proceedings of the IEEE conference on computer vision and
  pattern recognition}, pages 5505--5514, 2018.

\bibitem{yu2019free}
Jiahui Yu, Zhe Lin, Jimei Yang, Xiaohui Shen, Xin Lu, and Thomas~S Huang.
\newblock Free-form image inpainting with gated convolution.
\newblock In {\em Proceedings of the IEEE/CVF International Conference on
  Computer Vision}, pages 4471--4480, 2019.

\bibitem{zbontar2018fastmri}
Jure Zbontar, Florian Knoll, Anuroop Sriram, Tullie Murrell, Zhengnan Huang,
  Matthew~J Muckley, Aaron Defazio, Ruben Stern, Patricia Johnson, Mary Bruno,
  et~al.
\newblock {fastMRI: An open dataset and benchmarks for accelerated MRI}.
\newblock {\em arXiv preprint arXiv:1811.08839}, 2018.

\end{thebibliography}
}

\clearpage
\appendix

\renewcommand\thefigure{\thesection.\arabic{figure}}
\renewcommand{\thetable}{A.\arabic{table}}
\counterwithin{figure}{section}
\counterwithin{table}{section}
\renewcommand\thetheorem{\thesection.\arabic{theorem}}
\counterwithin{theorem}{section}
\counterwithin{lemma}{section}

\section*{\Large{\textbf{Supplementary Material}}}

\section{Mathematical Preliminaries}


\begin{definition}[Contraction on $\Rd^n$]
A function $\f:\Rd^n\mapsto \Rd^n$ is called a contraction mapping
if there exists a
real number 
$0\leq \lambda <1$  such that for all $\x$ and $\y$ in $\Rd^n$,
\begin{align}\label{eq:contraction}
\|\f(\x)-\f(\y)\|\leq \lambda \|\x-\y\|
\end{align}
\end{definition}
Using the intermediate value theorem for the function $\f(\x)$, we can easily see that
$\f(\x)$ is contracting with the rate $0\leq\lambda<1$ if $\f$ satisfies the following:
\begin{align}\label{eq:eigen}
\sigma_{\max}\left(\frac{\partial \f(\x)}{\partial \x}\right)\leq \lambda < 1
\end{align}
where $\sigma_{\max}(\Ab)$ denotes the largest singular value of a matrix $\Ab$.
Note that the contraction mapping in $\Rd^n$ is closely related to Lipschitz continuity,
and indeed the function that satisfies \eqref{eq:contraction} with any $\lambda>0$ is called
$\lambda$-Lipschitz continuous function.

Now, we provide a theorem for discrete stochastic contraction, which is
{slightly} modified from the contraction theorem of stochastic difference equation in \cite{pham2008analysis}.
\begin{theorem}\cite{pham2008analysis}
\label{thm:stc}
Consider the stochastic difference equation:
\begin{align}\label{eq:dsde}
\x_{i+1} = \f(\x_i,i)+g(\x_i,i)\w_{i}
\end{align}
where $\f(\cdot,i)$ is a $\Rd^n \mapsto\Rd^n$ function, 
$g(\cdot,i)$ is a $\Rd^n\times\Nd\mapsto\Rd$ {function} for each $i\in \Nd$, and $\{\w_i,i=1,2,\cdots\}$ is a sequence
of independent $n$-dimensional  zero mean unit variance Gaussian noise vectors.
Assume that the system satisfies the following two hypothesis:
\begin{itemize}
\item [(H1)] The function $\f(\cdot,i)$ is contracting with factor $\lambda$ in the sense of \eqref{eq:eigen} for all $i\in \Nd$.
\item [(H2)]
$\mathrm{Tr}(g(\x,i)\I g(\x,i)) \leq C, \quad \forall \x,i$.
\end{itemize}
Then, for two sample trajectory $\x_i$ and $\tilde\x_i$ that  satisfies \eqref{eq:dsde}, we have
\begin{align}
\Ed \|\x_i-\tilde\x_i\|^2 \leq \frac{2{C}}{1-\lambda^2}+{\lambda}^{2i}  \Ed \|\x_0-\tilde\x_0\|^2
\end{align}
\end{theorem}

The following corollary is a simple consequence of Theorem~\ref{thm:stc}.
\begin{corollary}
Consider the stochastic difference equation associated with the data fidelity term:
\begin{align}
\x'_{i+1} &= \f(\x_i,i)+\sigma(\x_i,i)\z_{i}\\
\x_{i+1} & = \A\x'_{i+1}+\bb \label{eq:const3}
\end{align}
where the $\A\in \Rd^{n\times n}$ is a non-expansive linear mapping,
and $\f(\x,i)$ and $\sigma(\x,i)$ satisfies (H1) and (H2).
Then, for two sample {trajectories} $\x_i$ and $\tilde\x_i$ that  satisfies \eqref{eq:dsde}, we have
\begin{align}
\Ed \|\x_i-\tilde\x_i\|^2 \leq \frac{2{C}\tau}{1-\lambda^2}+{(\lambda)}^{2i}  \Ed \|\x_0-\tilde\x_0\|^2
\end{align}
where $\tau = \frac{\mathrm{Tr}(\A^T\A)}{n}$.
\end{corollary}
\begin{proof}
After the application of \eqref{eq:const3}, we have
\begin{align*}
\x_{i+1} & =\underbrace{\A\f(\x_i,i)+ \bb}_{\tilde \f(\x_i,i)} + {\sigma(\x_i,i)\A {\z_i}}
\end{align*}
Therefore, we have
\begin{align*}
\sigma_{\max}\left( \frac{\partial \tilde \f(\x,i)}{\partial \x}\right) &\leq \sigma_{\max}(\A)\sigma_{\max}\left( \frac{\partial \f(\x,i)}{\partial \x}\right) \\
&=  \lambda
\end{align*}
as $ \sigma_{\max}(\A)\leq 1$ for a non-expansive linear mapping.
Furthermore, we have
\begin{align*}
\mathrm{Tr}(g(\x,i)\A^T\A g(\x,i) ) &= g(\x,i)^2 \mathrm{Tr}(\A^T\A)\\
&= \frac{\mathrm{Tr}(\A^T\A)}{n}C = C\tau
\end{align*}
Therefore, we have
\begin{align}
\Ed \|\x_i-\tilde\x_i\|^2 \leq \frac{2C\tau}{1-\lambda^2}+{\lambda}^{2i}  \Ed \|\x_0-\tilde\x_0\|^2
\end{align}
\end{proof}
%
%


\begin{lemma}
\label{lem:dsm}
Let $s_{\theta}(\x_i, i)$ be a sufficiently expressive parameterized score function so that
\begin{align}\label{eq:s}
s_{\theta}(\x_i, t) =  \frac{\partial}{\partial\x_i} \log p_{0i}(\x_i|\x_0)
\end{align}
Then, we have 
\begin{equation}
    \frac{\partial}{\partial\x_i}\s_{\theta}(\x_i, t) = -\frac{1}{b_i^2}\I.
\end{equation}
where
\begin{align}
b_i^2 = \begin{cases}  {{1-\bar\alpha_i}} , & \mbox{(DDPM)} \\  \sigma_i^2-\sigma_0^2,& \mbox{(SMLD)} \end{cases}
\end{align}
\end{lemma}
\begin{proof}
The forward diffusion is given by
\begin{align}
\x_{i} = a_i{\x}_0 + b_i\z
\end{align}
{where $\z\sim \Nc(0,\I)$ and  $(a_i,b_i)$ are defined in \eqref{eq:DDPMf} and \eqref{eq:SMLDf} for DDPM and SMLD, respectively.}
Using \eqref{eq:s}, we have
\begin{align}
&    \frac{\partial}{\partial\x_i}\left(\s_{\theta^*}(\x_i, i)\right)^T\\
 &= 
    \frac{\partial}{\partial\x_i}\left(\frac{\partial}{\partial \x_i }\log p_{0i}(\x_i|\x_0)\right)^T \\ 
    &=
        \frac{\partial}{\partial\x_i}\left(\frac{\partial}{\partial \x_i }\Big(-\frac{\|\x_i - a_i\x_0\|^2}{2b_i^2}\Big)  \right)^T \\ 
    &=
    \frac{\partial}{\partial\x_i}\Big(-\frac{\x_i - a_i\x_i}{b_i^2}\Big)^T \\ &=
    -\frac{1}{b_i^2}\I, 
\end{align}
where $^T$ denotes the transpose. This concludes the proof.
\end{proof}

\section{Proof of Theorem~\ref{thm:contraction}}

Let $N$ be the standard reverse diffusion step when starting from $T=1$. Then, 
the number of discretization step for our method is given $N'=Nt_0 < N$ so that
$t_0$ can refer to  the acceleration factor. 
We further define a new index $i = N'-j$ to convert the reverse diffusion index $j=N',\cdots,1$
to a forward direction index $i=0,1,\cdots, N'$.
This does not change the contraction property of the stochastic difference equation.
Therefore, without loss of generality, we use the aforementioned contraction property of stochastic
difference equation for the  index $i=0,1,\cdots, N'$.
Now, we are ready to provide the proof.

\subsection{DDPM}

In DDPM, the discrete version of the forward diffusion is given by Eq.~\eqref{eq:DDPMf}, and the reverse diffusion is given by eq.~\eqref{eq:DDPMr}.
Here, $\z_\theta(\x,i)$ is trained by
\begin{align}\label{eq:ztheta}
    \min_\theta \Ed_i \Ed_{\x(0)} \Ed_{\z \sim \Nc(\textbf{0}, \I)}\Big[
    \| \z - \z_\theta(\sqrt{\bar{\alpha}_i}\x(0) + \sqrt{1-\bar{\alpha}_i}\z, i) \|^2
    \Big].
\end{align}
It was shown that $\z_\theta(\x,i)$ is a scaled version of the score function \cite{song2020score}:
\begin{align}
\s_\theta(\x,i) = -\frac{1}{\sqrt{1-\bar\alpha_i}}\z_\theta(\x,i)
\end{align}
which leads to
\begin{align}
\x_{i-1}=\underbrace{\frac{1}{\sqrt{\alpha_i}}\Big(\x_i + {(1 - \alpha_i)}\s_\theta(\x_i, i)\Big)}_{\f(\x_i,i)} + \sigma_i \z,
\end{align}
Thus, we have
\begin{align*}
\frac{\partial \f^T(\x_i,i)}{\partial \x_i} &=  \frac{1}{\sqrt{\alpha_i}}\left(\I+(1-\alpha_i)\frac{\partial \s_\theta^T(\x_i,i)}{\partial \x_i}\right)\\
&= \frac{1}{\sqrt{\alpha_i}}\left(1- \frac{1-\alpha_i}{1-\bar\alpha_i}\right)\I\\
&= \frac{1}{\sqrt{\alpha_i}}\frac{\alpha_i-\bar\alpha_i}{1-\bar\alpha_i} \I\\
&=\sqrt{\alpha_i}\frac{1-\bar\alpha_{i-1}}{1-\bar\alpha_i} \I 
\end{align*}
Therefore, the contraction rate is given by
\begin{align}
\lambda=\max_{i\in[N']}\sqrt{\alpha_i}\left(\frac{1-\bar\alpha_{i-1}}{1-\bar\alpha_i}\right)<1
\end{align}
as $0<\alpha_i,\bar\alpha_i<1$.
Furthermore, we can easily show that 
$$C=n\max_{i\in[N']}({1 - \bar\alpha_i})=n(1-\bar\alpha_N),$$
as $\bar\alpha_i$ is decreasing with $i$.

\subsection{SMLD: Discrete Version of VE-SDE}

In discrete version of VE-SDE, the forward diffusion is given by \eqref{eq:SMLDf}. The associated reverse diffusion is given by \eqref{eq:SMLDr}.
Thus, we have
\begin{align*}
\frac{\partial \f^T(\x_i,i)}{\partial \x_i} &= \I+(\sigma_{i}^2 - \sigma_{i-1}^2)\frac{\partial \s_\theta^T(\x_i,i)}{\partial \x_i} \\
&= \left(1- \frac{\sigma_{i}^2 - \sigma_{i-1}^2}{\sigma_{i}^2 - \sigma_{0}^2}\right)\I\\
&=  \frac{\sigma_{i-1}^2 - \sigma_{0}^2}{\sigma_{i}^2 - \sigma_{0}^2} \I
\end{align*}
and the contraction rate is given by
\begin{align}
\lambda= \max_{i\in[N']}\frac{\sigma_{i-1}^2 - \sigma_{0}^2}{\sigma_{i}^2 - \sigma_{0}^2}<1
\end{align}
as $\sigma_i$ is increasing with $i$.
Furthermore, we can easily show that 
$${C=n\max_{i\in[N']}\sigma_i^2-\sigma_{i-1}^2}$$

\subsection{DDIM}
\label{sec:supp_ddim}

{The DDIM forward diffusion can be set identically to the forward diffusion of DDPM~\eqref{eq:DDPMf}, whereas the reverse diffusion is given as~\eqref{eq:DDIM}.
In fact, with a proper reparameterization, one can cast DDIM such that it is equivalent to the discrete version of VE-SDE without noise terms.}
More specifically, if we define the following reparametrization:
\begin{align}
\bar\x_{i}= \frac{\x_i}{\sqrt{\bar\alpha_i}}
\end{align}
then \eqref{eq:DDIM} becomes
\begin{align}
\bar\x_{i-1}&= \bar\x_{i}+\left(\sigma_{i-1}-\sigma_i\right)\z_\theta(\x_i,i)
\end{align}
where
\begin{align}
\sigma_i=\frac{\sqrt{1-\bar\alpha_{i}}}{\sqrt{\bar\alpha_{i}}}
\end{align}
Furthermore, the corresponding score function with respect to the reparameterization is
\begin{align}
\s_\theta(\bar\x_i,i)=-\frac{\z_\theta(\x_i,i)}{\sigma_i}
\end{align}
so that we have
\begin{align}
\bar\x_{i-1}&= \bar\x_{i}-\left(\sigma_{i-1}-\sigma_i\right)\sigma_i\s_\theta(\bar\x_i,i)
\end{align}
The forward diffusion \eqref{eq:DDPMf}
can be equivalently represented by the reparameterization as:
\begin{align}
\bar\x_i = \bar\x_0 +\sigma_i\zb
\end{align}
as $\alpha_0=1$.
Therefore, we have
\begin{align}
\frac{\partial \f^T}{\partial \bar \x_i}(\bar\x_i) =  \left(1+ \frac{\sigma_{i-1}-\sigma_i}{\sigma_i}\right)\I= \frac{\sigma_{i-1}}{\sigma_i}\I
\end{align}
and the contraction rate is given by
\begin{align}
\lambda= \max_{i\in[N']}\frac{\sigma_{i-1}}{\sigma_{i}}<1
\end{align}
as $\sigma_i$ is increasing with $i$.
Furthermore, we can easily show that  $C=0$ as there is no noise term.

\section{{Proof of Theorem~\ref{thm:sp}}}
\renewcommand{\rev}{\textcolor{red}}

For some of the proofs, we borrow more tight inequality to obtain the result.
In fact, the inequality of stochastic contraction
\begin{align}
\bar\varepsilon_{0,r} \leq \frac{2{C}\tau}{1-\lambda^2}+{\lambda}^{2N'} \bar\varepsilon_{N'}\label{eq:sc}
\end{align}
is a rough estimation of recursive inequality \cite{pham2008analysis}
\begin{align}
    \bar\varepsilon_{j-1,r} \le \lambda^2_j \bar\varepsilon_{j,r} + 2C_j \tau,
\end{align}
where $\bar\varepsilon_{j,r}$ denotes  the  estimation  error  between  reverse conditional  diffusion  path  down  to $j$.
Accordingly, we have
\begin{align}
    \bar\varepsilon_{0,r} \le \bar\varepsilon_{N,r} \prod_{j=0}^{N} \lambda^2_j + \sum_{j=1}^{N} \left(2C_j \tau \prod_{i=1}^{j-1}\lambda^2_i \right), \label{eq:full-ineq}
\end{align}
which is reduced to (\ref{eq:sc})
when $\lambda_j$ and $C_j$ are uniformly bounded by $\lambda$ and $C$, respectively.

Now, our proof strategy is as follows.
We specify reasonable conditions on $\{\beta_i\}$ or $\{\sigma^2_i\}$, which are satisfied by the existing DDPM, SLMD, and DDIM scheduling approaches.
Then, for any $0 < \mu \leq 1$,  our goal is to to show that there exists $N'$ such that
\[ \bar\varepsilon_{0,r} \le \mu \varepsilon_0,\]
and $N'$ decreases as $\varepsilon_0$ gets smaller.

\subsection{DDPM}
Without loss of generality, we assume that  ground truth image and the corrupted image are normalized within range $[0, 1]$, i.e. $\x, \bar\x \in [0,1]^n$.
Then, we have
\begin{align}\label{eq:norm}
\varepsilon_0 = \| \x - \tilde\x \|^2 \le n. 
\end{align}
We choose $N'$ such that
\begin{align}
   N' \beta_{N'} &\ge 2 \log \left(\frac{4n}{\mu\varepsilon_0}\right) \label{eq:ddpm-lower}\\
   N' \beta_{N'} &\le \frac{\mu \varepsilon_0}{4n \tau}. \label{eq:ddpm-upper}
\end{align}
We separately investigate each term in (\ref{eq:full-ineq}).
First,
from theorem 1,
\begin{align*}
  \bar\varepsilon_{N,r} &= a_{N'}^2\varepsilon + 2b_{N'}^2 n \\
  &= \bar\alpha_{N'} \varepsilon_0 + (1-\bar\alpha_{N'}) 2n \\
  &= 2n +\bar\alpha_{N'}(\varepsilon_0-2n)\\
  &\le 2n
\end{align*}
where the last inequality comes from \eqref{eq:norm}.
Subsequently,
\begin{align*}
    &\sum_{j=1}^{N'} \left(2C_j \tau \prod_{i=1}^{j-1}\lambda^2_i \right)\\
    &=\sum_{j=1}^{N'} \left(2n(1-\alpha_j) \tau \prod_{i=1}^{j-1}\lambda^2_i \right)\\
    &\le 2n \tau \sum_{j=1}^{N'} \beta_j \cdot 1\\
    &\le 2n \tau N' \beta_{N'} \le \frac{\mu\varepsilon_0}{2}.
\end{align*}
where the first inequality comes from $\prod_{i=1}^{j-1} \lambda_i^2 \le \prod_{i=1}^{j-1} 1 \le 1$ and the last equality is from (\ref{eq:ddpm-upper}).
Therefore,
\begin{align}
    \bar\varepsilon_{0,r} &\le \bar\varepsilon_{N',r} \prod_{j=0}^{N'} \lambda^2_j + \sum_{j=1}^{N'} \left(2C_j \tau \prod_{i=1}^{j-1}\lambda^2_i \right) \notag\\
    & \le 
    2n \cdot e^{-\frac{N'\beta_{N'}}{2}} + \frac{\mu\varepsilon_0}{2} \label{eq:lambda-bound}\\
    &\le 2n \cdot \frac{\mu\varepsilon_0}{4n} + \frac{\mu \varepsilon_0}{2} \le \mu \varepsilon_0 , \notag
\end{align}
where the third inequality holds by (\ref{eq:ddpm-lower}), 
and the inequality in (\ref{eq:lambda-bound}) comes from Lemma~\ref{lem:exp} (see below).
Furthermore, from (\ref{eq:ddpm-upper}), we can
see that $N'$ becomes smaller for a smaller $\varepsilon_0$.
This concludes the proof of DDPM.

\begin{lemma}\label{lem:exp}
    $$\prod_{j=1}^{N'} \lambda^2_j \le e^{-\frac{N'\beta_{N'}}{2}}.$$
\end{lemma}
\begin{proof} [Proof of Lemma~\ref{lem:exp}]
\begin{align*}
    \prod_{j=1}^{N'} \lambda^2_j &= \prod_{j=1}^{N'} \alpha_j \cdot \frac{(1-\bar\alpha_{j-1})^2}{(1-\bar\alpha_{j})^2} \\
    &\le \prod_{j=1}^{N'} \alpha_j \\
    &\le \left( \frac{1}{N'} \sum_{j=1}^{N'} \alpha_j \right)^{N'}\\
    &= \left( 1 - \frac{1}{N'} \sum_{j=1}^{N'} \beta_j \right)^{N'}\\
    &= \left(  1- \frac{\beta_{N'}}{2} \right)^{N'}
\end{align*}
where the first inequality comes from $\bar\alpha_{j}=\bar\alpha_{j-1}\alpha_j \le \bar\alpha_{j-1}$,
and the second inequality is the inequality of arithmetic and geometric means,
 and the third equality is from the linear increasing $\beta_j$ from $\beta_0=0$.
Finally, using
\begin{equation}
    e^x \ge \left(1+\frac{x}{N}\right)^N \quad \text{for } N\ge 1, |x|\le N \label{eq:exponential}
\end{equation}
we have
\begin{align*}
 \prod_{j=1}^{N'} \lambda^2_j &\le e^{-\frac{N\beta_{N'}}{2}},
\end{align*}
This concludes the proof.
\end{proof}

\subsection{SMLD}

Assume that the minimum and maximum values of variance satisfy the following:
\begin{align}
    \sigma^2_{\min} &< \frac{\mu^{\frac{3}{2}}\varepsilon_0}{8n} \label{eq:sigmin}\\
    \sigma^2_{\max} &> \frac{\mu\varepsilon_0}{4n} \ .
    \label{eq:sigmax} 
\end{align}
Then, using \eqref{eq:sigmin},
\begin{align*}
    \log\left(\frac{2}{\sqrt{\mu}}\right) &< \log\left(\frac{\mu\varepsilon_0}{4n\sigma^2_{\min}}\right),
\end{align*}
and thus
\begin{align}
\label{eq:ineq1}
    \frac{\log(2/\sqrt{\mu})}{\log(\sigma^2_{\max}/\sigma^2_{\min})} < \frac{\log(\mu\varepsilon_0/4n\sigma^2_{\min})}{\log(\sigma^2_{\max}/\sigma^2_{\min})}.
\end{align}
In addition, from \eqref{eq:sigmax}, we have
\begin{align*}
    \frac{\mu\varepsilon_0}{4n\sigma^2_{\min}} < \frac{\sigma^2_{\max}}{\sigma^2_{\min}},
\end{align*}
and hence
\begin{align}
\label{eq:ineq2}
    \frac{\log(\mu\varepsilon_0/4n\sigma^2_{\min})}{\log(\sigma^2_{\max}/\sigma^2_{\min})} < 1.
\end{align}
Combining \eqref{eq:ineq1} with \eqref{eq:ineq2}, we arrive at
\begin{align}
\label{eq:ineq_c}
    \frac{\log(2/\sqrt{\mu})}{\log(\sigma^2_{\max}/\sigma^2_{\min})} < \frac{\log(\mu\varepsilon_0/4n\sigma^2_{\min})}{\log(\sigma^2_{\max}/\sigma^2_{\min})} < 1.
\end{align}
Now, we can choose $N'$ such that it satisfies the following conditions:
\begin{align}
\begin{split}
    \frac{N'-1}{N-1} &\geq \frac{\log(2/\sqrt{\mu})}{\log(\sigma^2_{\max}/\sigma^2_{\min})} \\
    \frac{N'-1}{N-1} &\leq \frac{\log(\mu\varepsilon_0/4n\sigma^2_{\min})}{\log(\sigma^2_{\max} / \sigma^2_{\min})} \label{eq:up}
\end{split}
\end{align}
This leads to the following bounds
\begin{align}
\begin{split}
    \left(\frac{\sigma^2_{\max}}{\sigma^2_{\min}}\right)^\frac{N'-1}{N-1} &\geq \frac{2}{\sqrt{\mu}}\\
    n\sigma^2_{\min}\left(\frac{\sigma^2_{\max}}{\sigma^2_{\min}}\right)^\frac{N'-1}{N-1} &\leq \frac{\mu\varepsilon_0}{4}.
\end{split}\label{eq:smld_bound}
\end{align}
On the other hand, in the geometric scheduling of noise, for all $i$, we have
\begin{align}
    \lambda &= \sigma^2_{\min}\left(\frac{\sigma^2_{\max}}{\sigma^2_{\min}}\right)^{\frac{i-1}{N-1}}\bigg/\sigma^2_{\min}\left(\frac{\sigma^2_{\max}}{\sigma^2_{\min}}\right)^{\frac{i-2}{N-1}}\\
    &= \left(\frac{\sigma^2_{\min}}{\sigma^2_{\max}}\right)^{\frac{1}{N-1}} \text{ and} \notag \\
    C &= n \max\sigma_i^2 \left( 1- \frac{\sigma_{i-1}^2}{\sigma_i^2}\right) = n \sigma^2_{N'}(1-\lambda). \label{eq:C1}
\end{align}
where
$$\sigma_N' = \sigma_{\text{min}}\left(\frac{\sigma_{\text{max}}}{\sigma_\text{min}}\right)^{\frac{N'-1}{N-1}}.$$
Note that from \eqref{eq:smld_bound},
\begin{align}
    2n\sigma^2_{N'} = 2n\sigma^2_{\min}\left(\frac{\sigma^2_{\max}}{\sigma^2_{\min}}\right)^\frac{N'-1}{N-1} \leq \frac{\mu\varepsilon_0}{2},
\label{eq:smld_bound_f1}
\end{align}
and
\begin{align}
    \left(\frac{\sigma^2_{\min}}{\sigma^2_{\max}}\right)^\frac{2(N'-1)}{N-1} \leq \frac{\mu}{4}.
\label{eq:smld_bound_f2}
\end{align}
Hence, by plugging in  \eqref{eq:C1} to \eqref{eq:sc}, we have
\begin{align*}
    \bar\varepsilon_{0,r} &\leq \frac{2C\tau}{1 - \lambda^2} + \lambda^{2N'}\bar\varepsilon_{N'}\\
    &= \frac{2n\sigma^2_N(1-\lambda)\tau}{(1+\lambda)(1-\lambda)} + \left(\frac{\sigma^2_{\min}}{\sigma^2_{\max}}\right)^\frac{2N'}{N-1}(\varepsilon_0 + 2n\sigma^2_{N'})\\
    &\leq 2n\sigma^2_{N'}\frac{\tau}{1+\lambda} + \left(\frac{\sigma^2_{\min}}{\sigma^2_{\max}}\right)^\frac{2(N'-1)}{N-1}(\varepsilon_0 + 2n\sigma^2_{N'})\\
    &\leq \frac{\mu\varepsilon_0}{2} + \frac{\mu}{4}\left(\varepsilon_0 + \frac{\mu\varepsilon_0}{2}\right)\\
    &\leq \frac{\mu\varepsilon_0}{2} + \frac{\mu}{4}(\varepsilon_0 + \varepsilon_0)\\
    &= \mu\varepsilon_0,
\end{align*}
where the third inequality comes from the bounds in \eqref{eq:smld_bound_f1}, \eqref{eq:smld_bound_f2}, and the fact that $\tau = \frac{tr(A^T A)}{n} <1$ for a non-expansive linear mapping $A$. 

Finally, we can easily see that the value $N'$ satisfying (\ref{eq:up}) decreases as $\varepsilon_0$ decreases.

\subsection{DDIM}

In DDIM, we have $C_j = 0$ for Eq.~(\ref{eq:full-ineq}).
Let $\sigma_0$ and $N'$ satisfy the following:
\begin{align}
\sigma_0^2  &\le \frac{\mu\varepsilon_0}{4n}\label{eq:sigma0} \\
\sigma^2_{N'}& \ge \frac{\varepsilon_0}{2n} \label{eq:NDDIM}
\end{align}
Then, we have
\begin{align*}
    \bar\varepsilon_0,r &\le \bar\varepsilon_{N,r} \prod_{j=1}^{N} \lambda_j^2 \\
    &\le (\varepsilon_0 + \sigma_{N'}^2 2n) \cdot \frac{\sigma_0^2}{\sigma_{N'}^2} \\
    & \le \mu \varepsilon_0
\end{align*}
where the second equality comes from $\lambda_j=\sigma_{j-1}/\sigma_j$ and
the last equality comes from Eqs.~(\ref{eq:sigma0}) and (\ref{eq:NDDIM}).

We can also easily see that the minimum value $N'$ satisfying (\ref{eq:NDDIM}) decreases as $\varepsilon_0$ decreases,
as $\sigma_i^2$ is an increasing sequence in DDIM.

\section{Implementation detail}
\label{sec:supp_algorithm}

In this section, we provide detailed explanation of discrete version of CCDF for each application.
Again, the number of discretization step for our method is given $N'=Nt_0 < N$ where
$t_0$ refers to the acceleration factor.

\subsection{Super-resolution and Image Inpainting}

For these problems,
 we employ the discretized version of the VP-SDE, which has shown impressive results on conditional generation~\cite{choi2021ilvr, dhariwal2021diffusion}. Namely, we use DDPM~\cite{ho2020denoising}, with several strategies introduced in improved DDPM (IDDPM)~\cite{nichol2021improved} for both training the score function and for reverse diffusion procedure.
 
 The modified reverse diffusion is given by
 \begin{align}
{\x'_{i-1}=\frac{1}{\sqrt{\alpha_i}}\Big(\x_i + (1-\alpha_i) \s_\theta(\x_i, i)\Big) +\sqrt{\sigma_i}\z,}
\end{align}
where $\sigma_i$ is given by
\begin{equation}\label{eq:v}
    \sigma_i = \exp(v\log\beta_i + (1 - v)\log\Tilde{\beta}_i),
\end{equation}
letting model variance to be learnable in a range $[\beta_i, \Tilde{\beta}_i]$,
where $\tilde\beta_i$ {is given by $\tilde\beta_i = \frac{1 - \bar\alpha_{i-1}}{1 - \bar\alpha_i}\beta_i$.}
In \eqref{eq:v}, $v$ is the learnable parameter so that 
it can be trained using the variational lower-bound penalty
introduced in \cite{ho2020denoising, nichol2021improved}.


Specifically, $v$ and the score function $\s_\theta$ are trained using the following objective
\begin{equation}
\label{eq:vp_train}
    L_{total}(\theta,v) = L_{simple}(\theta) + \lambda L_{VLB}(v),
\end{equation}
where $L_{simple}(\theta)$ is given in \eqref{eq:ztheta} and
we apply stop-gradient for the $L_{VLB}$ so that the gradient of the loss contributes only to estimating the model variance. 

{For the training of score function, we use a U-Net  architecture as used in~\cite{nichol2021improved} with the loss function as given in \eqref{eq:vp_train}. Multi-headed attention~\cite{vaswani2017attention} was used only at the 16$\times$16 resolution. Linear beta noise scheduling~\cite{ho2020denoising} with $\beta_{\text{min}}=0.0001$ and $\beta_{\text{max}}=0.02$ were used, with $N=1000$ discretization. We train the model with a batch size of 2, and a static learning rate of \code{1e-4} with Adam~\cite{kingma2014adam} optimizer for 5M steps. Exponential moving average (EMA) rate of 0.9999 was applied to the model.}

For super-resolution, we define a blur kernel $\hb_D$ which is defined by  successive applications of the downsampling filter by
a factor $D$, and upsampling filter by a factor $D$.
This can be represented as a matrix multiplication:
\begin{align}
\Pb \x' :=\hb_D \ast \x'.
\end{align}
where $\x'$ denotes intermediate estimate from the reverse diffusion.
Then, we use the following data consistency iteration:
\begin{align}
\x_i=  {(\Ib-\Pb)\x_i'}  + {{\hat{\x}_i}},
\end{align}
{where $\x_i$ is the current estimate, and
$\hat \x_i$ is the forward propagated image from the initial measurement $\hat \x(0)$:}
\begin{align}
\hat\x_{i} = \sqrt{\bar{\alpha}_{i}}\hat{\x}_0 + \sqrt{1 - \bar{\alpha}_{i}}\z
\end{align}
Therefore, we have
$$\Ab= \I-\Pb,\quad \bb={{\hat{\x}_i}}.$$
We can easily see that $\sigma_{\max}(\Ab)\leq 1$ for the normalized filter $\hb_D$.

Similarly, for the case of image inpainting,  $\P$ is just a diagonal matrix with 1 at the measured locations
and 0 on the unmeasured locations so that  $\sigma_{\max}(\Ab)\leq 1$.

The resulting pseudo-code implementation of the algorithm is
given in Algorithm~\ref{alg:SR_inpaint}.

\begin{algorithm}[!hbt]
\caption{Accelerated Super-resolution / inpainting (VP, Markov)}
\begin{algorithmic}[1]
\Require {${\x_0}, \hat{\x}_0, N', \{\alpha_i\}_{i=1}^{N'}, \{\sigma_i\}_{i=1}^{N'}, \s_{\theta}$}
\State $\z \sim \Nc(\textbf{0}, \I)$
\State {$\x_{N'} \gets \sqrt{\bar{\alpha}_{N'}}{\x_0} + \sqrt{1 - \bar{\alpha}_{N'}}\z$} \Comment{Forward diffusion}
\For{$i = N'$ to $1$} \Comment{Reverse diffusion} \do \\
\State $\x'_{i-1} \gets \frac{1}{\sqrt{\alpha_i}}(\x_i +{(1 - \alpha_i)}\s_\theta(\x_{i}, i))$
\State $\z \sim \Nc(\textbf{0}, \I)$
\State $\x_{i-1} \gets \x'_{i-1} + \sigma_i\z$ \Comment{Unconditional update}
\State $\z \sim \Nc(\textbf{0}, \I)$
\State $\hat{\x_i} \gets \sqrt{\bar{\alpha}_i}\hat{\x}_0 + \sqrt{1 - \bar{\alpha}_i}\z$
\State $\x_{i-1} = (\I - \P) \x_{i-1} + {\hat{\x}_i}$
    \NoNumber{\Comment{Measurement consistency}}
\EndFor
\State \textbf{return} {$\x_0$}
\end{algorithmic}\label{alg:SR_inpaint}
\end{algorithm}

\subsection{{DDIM for Super-resolution/Inpainting}}

{Note that we can use the same score function trained for DDPM, and use it in DDIM sampling~\cite{song2020denoising}. Here, we study the effect on combining DDIM together with the proposed method to achieve even further acceleration. All we need to do is modify the unconditional update step, arriving at Algorithm~\ref{alg:SR_inpaint_ddim}.}

\begin{algorithm}[!hbt]
\caption{Accelerated Super-resolution / inpainting (VP, markov) + DDIM}
\begin{algorithmic}[1]
\Require {${\x_0}, \hat{\x}_0, N', \{\alpha_i\}_{i=1}^{N'}, \{\sigma_i\}_{i=1}^{N'}, \s_{\theta}$}
\State $\z \sim \Nc(\textbf{0}, \I)$
\State {$\x_i \gets \sqrt{\bar{\alpha}_i}{\x_0} + \sqrt{1 - \bar{\alpha}_i}\z$} \Comment{Forward diffusion}
\For{$i = N_0$ to $1$} \Comment{Reverse diffusion} \do \\
\State \begin{varwidth}[t]{\linewidth}
$\x_{i-1} \gets \frac{1}{\sqrt{\alpha_i}}\x_i +$ \par
\hskip\algorithmicindent ${\Big(\frac{1 - \bar\alpha_i}{\sqrt{\alpha_i}} - \sqrt{(1 - \bar\alpha_i)(1 - \bar\alpha_{i-1})}\Big)\s_\theta(\x_{i}, i)}$
\end{varwidth}
\NoNumber{\Comment{Unconditional update}}
\State {$\z \sim \Nc(\textbf{0}, \I)$}
\State $\hat{\x_i} \gets \sqrt{\bar{\alpha}_i}\hat{\x}_0 + \sqrt{1 - \bar{\alpha}_i}\z$
\State ${\x_{i-1} = (\I - \P) \x_{i-1} + \hat{\x_i}}$
    \NoNumber{\Comment{Measurement consistency}}
\EndFor
\State \textbf{return} {$\x_0$}
\end{algorithmic}\label{alg:SR_inpaint_ddim}
\end{algorithm}

\subsection{MRI reconstruction}

For the task of MRI reconstruction, we ground our work on Score-MRI~\cite{chung2021score}, and modify the previous algorithm for our purpose. The algorithm is given in Algorithm \ref{alg:mri_recon}. Specifically, we use variance exploding (VE-SDE) with predictor-corrector (PC) sampling which gives optimal results for MR reconstruction. 
For the step size of the corrector (Langevin dynamics) step, we use the following
\begin{equation}
    \epsilon_i = 2r\frac{\|\z\|_2}{\|\s_\theta(\x_i, \sigma_i)\|_2},
\end{equation}
with $r = 0.16$ set as constant. 
For training the score function, we use the following minimization strategy:
\begin{align}\label{eq:score_cost_VESDE}
    \min_{\thetab} &\Ed_{t \sim U(\eta, 1)} \Ed_{\xb(0) \sim p_0} \Ed_{\xb(t) \sim \Nc(\xb(0),\sigma^2(t)\Ib)}\Big[ \\ \notag
    &\norm{\sigma(t)\sb_{\thetab} (\xb(t), t) - \frac{\xb(t) - \xb(0)}{\sigma(t)}}_2^2
    \Big],
\end{align}
with $\eta =$ \code{1e-5}, and
\begin{equation}
    \sigma(t) = \sigma_{\min} \left(\frac{\sigma_{max}}{\sigma_{\min}}\right)^t,
\end{equation}
with $\sigma_{\min} = 0.01, \sigma_{max} = 378$. {We construct a modified U-Net model introduced in~\cite{song2020score}, namely \code{ncsnpp}. Adam optimizer is used for optimization, with a static learning rate of \code{2e-4} for 5M steps. EMA rate of 0.999 is used, and gradient clipping is applied with the maximum value of 1.0.}

In compressed sensing MRI, the subsampled $k$-space data $\yb$ is obtained from underlying image $\xb$ as:
\begin{align}
\yb =\D \F \x
\end{align}
where $\F$ denote the Fourier transform and its inverse,
 $\D$ is a diagonal matrix indicating the $k$-space sampling
location and $\yb$ is the original zero-filled $k$-space data.

The associated  data consistency imposing operator is then defined by
\begin{align}
\x_i = (\I- \F^{-1}\D\F)\x_i'+\F^{-1}\D\yb
\end{align}
where  $\F^{-1}$ is the inverse Fourier transform.
Therefore, we have
$$\Ab= \I- \F^{-1}\D\F,\quad \bb=\F^{-1}\D\yb.$$
Again, we can easily see $\sigma_{\max}(\Ab)\leq 1$ as the Fourier
transform is orthonormal.

The corresponding pseudo-code implementation is shown in
Algorithm~\ref{alg:mri_recon}.

\begin{algorithm}[!hbt]
\caption{Accelerated MR reconstruction (VE, PC)}
\begin{algorithmic}[1]
\Require {${\x_0}, \y, N', \{\sigma_i\}_{i=1}^{N'}, \{\epsilon_i\}_{i=1}^{N'}, \s_{\theta}$}
\State $\z \sim \Nc(\textbf{0}, \I)$
\State $\x_{N'} \gets {\x_0} + \sigma_{N'}\z$ \Comment{Forward diffusion}
\For{$i = N'$ to $1$} \Comment{Reverse diffusion} \do \\
\State $\x'_{i-1} \gets \x_i + (\sigma_{i}^2 - \sigma_{i-1}^2)\s_\theta(\x_{i}, \sigma_{i})$
\State $\z \sim \Nc(\textbf{0}, \I)$
\State $\x_{i-1} \gets \x'_{i-1} + \sqrt{\sigma_{i}^2 - \sigma_{i-1}^2}\z$ \Comment{Predictor}
\State $\x_{i-1} = (\I- \F^{-1}\D\F)\x_i+\F^{-1}\D\yb$
    \NoNumber{\Comment{Measurement consistency}}
\State $\x'_{i-1} \gets \x_{i-1} + \epsilon_i\s_\theta(\x_{i}, \sigma_{i})$
\State $\z \sim \Nc(\textbf{0}, \I)$
\State $\x_{i-1} \gets \x'_{i-1} + \sqrt{2\epsilon_i}\z$ \Comment{Corrector}
\State $\x_{i-1} = (\I- \F^{-1}\D\F)\x_i+\F^{-1}\D\yb$
    \NoNumber{\Comment{Measurement consistency}}
\EndFor
\State \textbf{return} ${\x_0}$
\end{algorithmic}\label{alg:mri_recon}
\end{algorithm}

{All training and inference algorithms were implemented in PyTorch, and were performed on a single RTX 3090 GPU.}

\section{Additional Experiments}
\label{sec:supp_ae}

\subsection{Super-resolution}

\begin{figure}[!hbt]
    \centering\includegraphics[width=8cm]{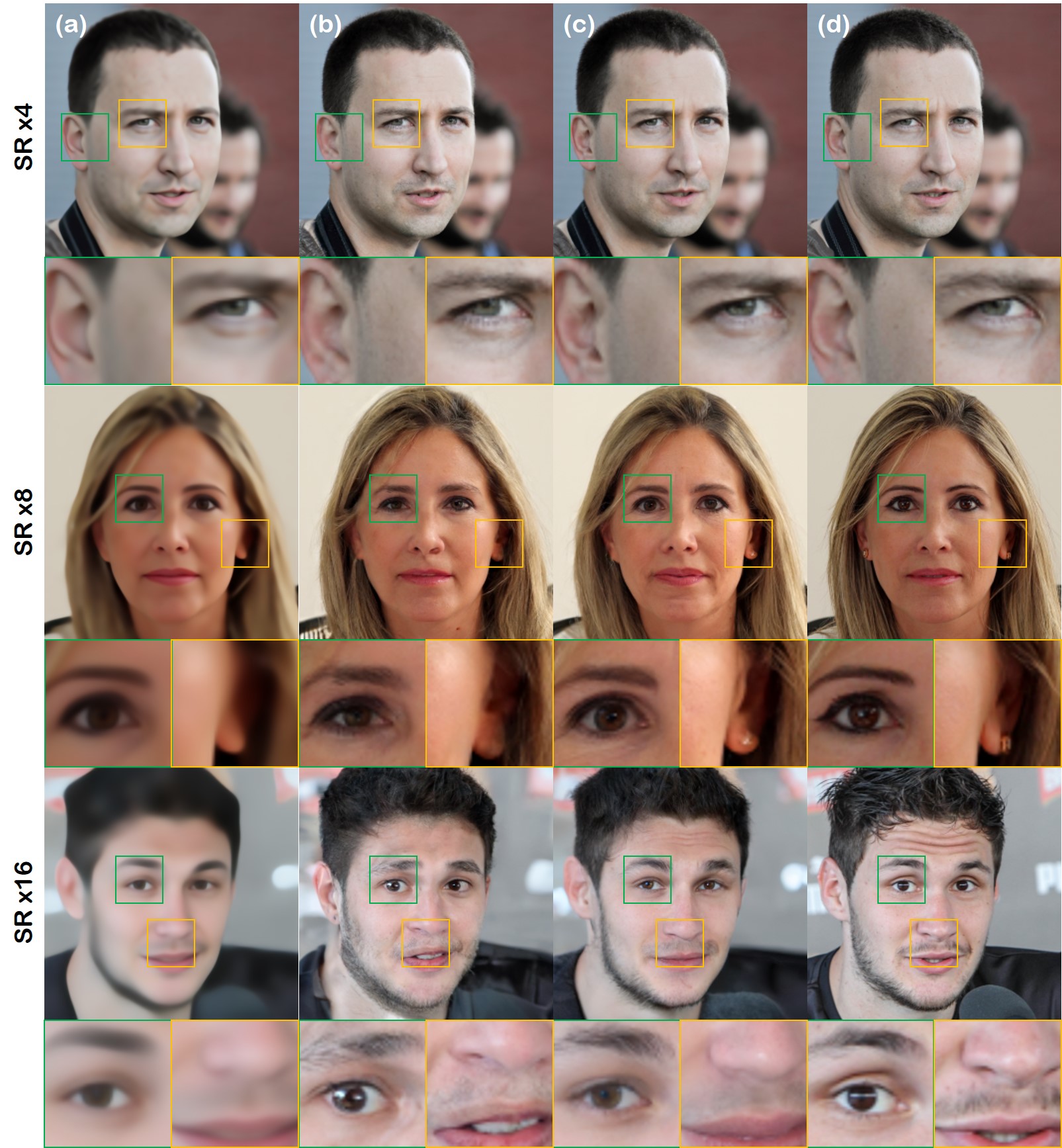}
    \caption{Comparison on SR task ($\times$4, $\times$8, $\times$16 for the 1$^{st}$, 2$^{nd}$, and 3$^{rd}$ row): (a) ESRGAN~\cite{wang2018esrgan},  (b) ILVR~\cite{choi2021ilvr} (1000 steps), (c) proposed method (100, 200, 300 steps for $\times$4, $\times$8, $\times$16 SR), (d) Ground Truth}
	\label{fig:SR_comparison}
\end{figure}

\noindent
\textbf{{Comparison study.}}~{In Fig.~\ref{fig:SR_comparison}, we compare the results of super-resolution using fairly large number of diffusion steps, as opposed to using only 20 number of diffusion steps as shown in Fig.~\ref{fig:SR_results}. This is a region where ILVR is known to perform well, as opposed to the few-step setting. While in Fig.~\ref{fig:SR_comparison}, ILVR uses 1000 steps of diffusion, the proposed method only uses 100, 200, and 300 steps of diffusion for $\times 4, \times 8,$ and $\times 16$, respectively. Nevertheless, the quality of reconstruction does not degrade, thanks to the contraction property of CCDF.}

\begin{figure}[!h]
    \centering\includegraphics[width=8cm]{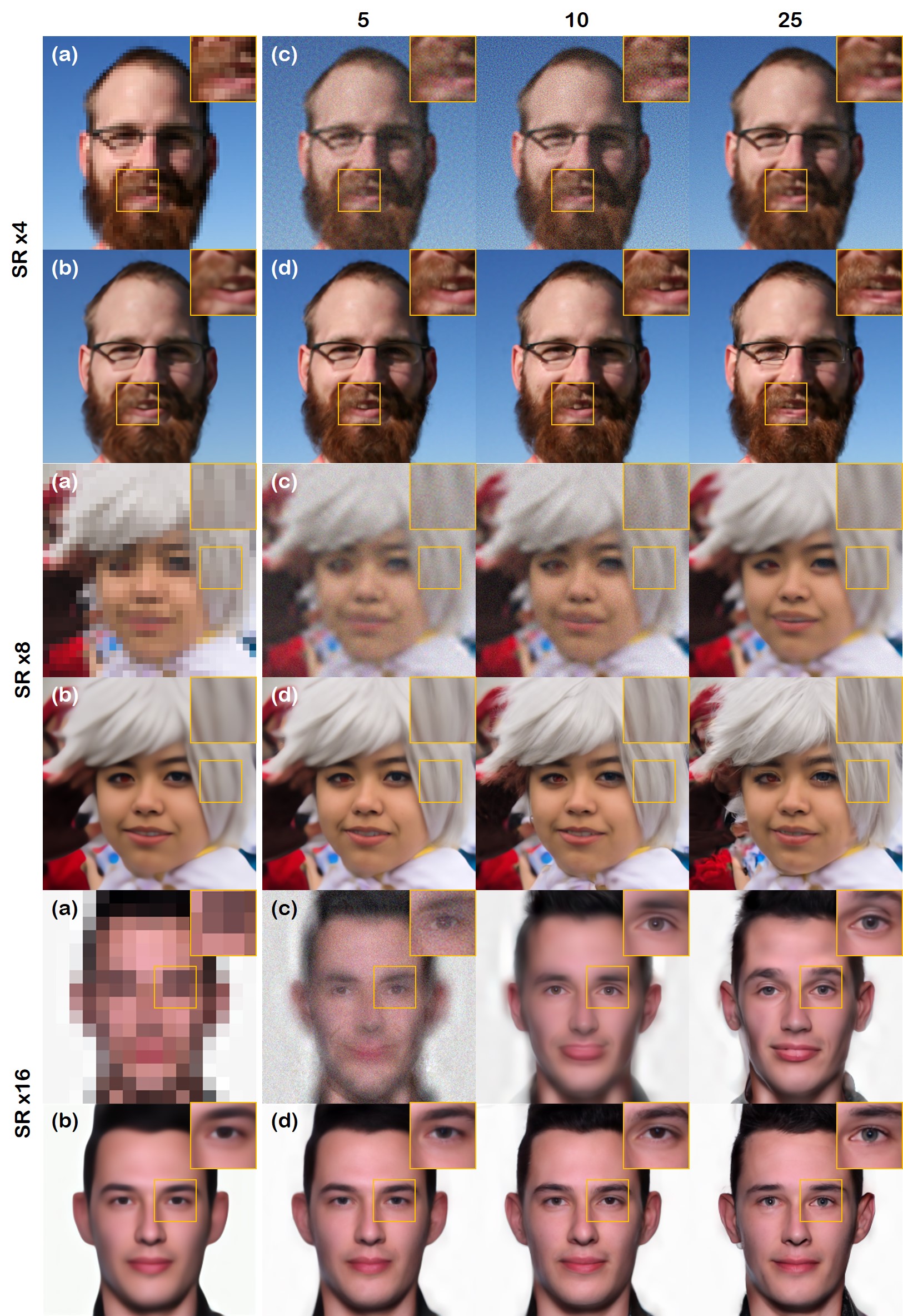}
    \caption{Results on SR task of FFHQ dataset with proposed method + DDIM. Numbers on top indicate number of reverse diffusion iterations. (a) LR image, (b) ground truth, (c) ILVR + DDIM, (d) CCDF + DDIM.}
	\label{fig:ddim_ffhq}
\end{figure}

\begin{table}[!hbt]
    \centering
    \resizebox{0.4\textwidth}{!}{
    \begin{tabular}{c|c|c|cccc}
    \hline
         & SR & method & 5 & 10 & 25 & 50 \\ \hline\hline
    \multirow{6}{*}{FFHQ} & \multirow{2}{*}{$\times 4$} & \thead{ILVR\\ +DDIM} & 120.53 & 114.61 & 87.15 & 81.85 \\
    & & \thead{proposed\\+DDIM} & \underline{72.34} & \textbf{69.39} & 78.83 & 82.72 \\ \cline{2-7}
    & \multirow{2}{*}{$\times 8$} & \thead{ILVR\\ +DDIM} & 147.44 & 115.30 & 101.37 & 93.72 \\
    & & \thead{proposed\\+DDIM} & 91.84 & \textbf{85.43} & \underline{87.43} & 94.89 \\ \cline{2-7}
    & \multirow{2}{*}{$\times 16$} & \thead{ILVR\\ +DDIM} & 147.44 & 115.30 & 101.37 & 93.72 \\
    & & \thead{proposed\\+DDIM} & 91.84 & \textbf{85.43} & \underline{87.43} & 94.89 \\ \hline
    \multirow{6}{*}{AFHQ} & \multirow{2}{*}{$\times 4$} & \thead{ILVR\\ +DDIM} & 63.79 & 55.57 & 40.22 & 30.57 \\
    & & \thead{proposed\\+DDIM} & \underline{17.57} & \textbf{17.19} & 20.87 & 30.22 \\ \cline{2-7}
    & \multirow{2}{*}{$\times 8$} & \thead{ILVR\\ +DDIM} & 106.94 & 67.06 & 51.75 & 45.96 \\
    & & \thead{proposed\\+DDIM} & \underline{35.03} & \textbf{31.70} & 35.62 & 45.17 \\ \cline{2-7}
    & \multirow{2}{*}{$\times 16$} & \thead{ILVR\\ +DDIM} & 163.98 & 94.68 & 69.60 & 65.33 \\
    & & \thead{proposed\\+DDIM} & 70.02 & \underline{59.01} & \textbf{49.36} & 64.61 \\ \hline
    \end{tabular}
    }
    \caption{FID($\downarrow$) scores on FFHQ and AFHQ test set for SR task with DDIM by varying the number of iterations.}
    \label{tab:FID_SR_ddim}
\end{table}

\noindent
\textbf{{Incorporation of DDIM.}}~{We provide additional SR results using CCDF + DDIM. In Figure~\ref{fig:ddim_ffhq}, we show an experiment with the FFHQ dataset, where we compare the combination of ILVR + DDIM, and proposed method + DDIM. For ILVR + DDIM, in order to reduce the number of iterations, we choose larger discretization steps used in DDIM. For the proposed method, we fix $N = 50$, and reduce the value of $t_0$ to achieve less iterations. In the figure, we confirm that our method can be used together with DDIM to create high-fidelity samples with as small as 5 reverse diffusion iterations, even when it comes down to extreme cases of SR $\times 8$ or SR $\times 16$. Additionally, we observe that the results with $t_0 \leq 0.5$ is {\em superior} to the $t_0 = 1.0$ counterparts, again confirming our theory.}

{The same trend can also be seen via quantitative metrics in Table~\ref{tab:FID_SR_ddim}. Using limited number of diffusion steps, FID score in the case of ILVR+DDIM grows exponentially, as we decrease the number of steps taken. Contrarily, our method is able to {\em improve} the metric by quite a margin, as opposed to using full diffusion with 50 steps in total. This trend is indeed similar to the experiments performed with DDPM.}

\begin{figure*}[!hbt]
    \centering\includegraphics[width=14cm]{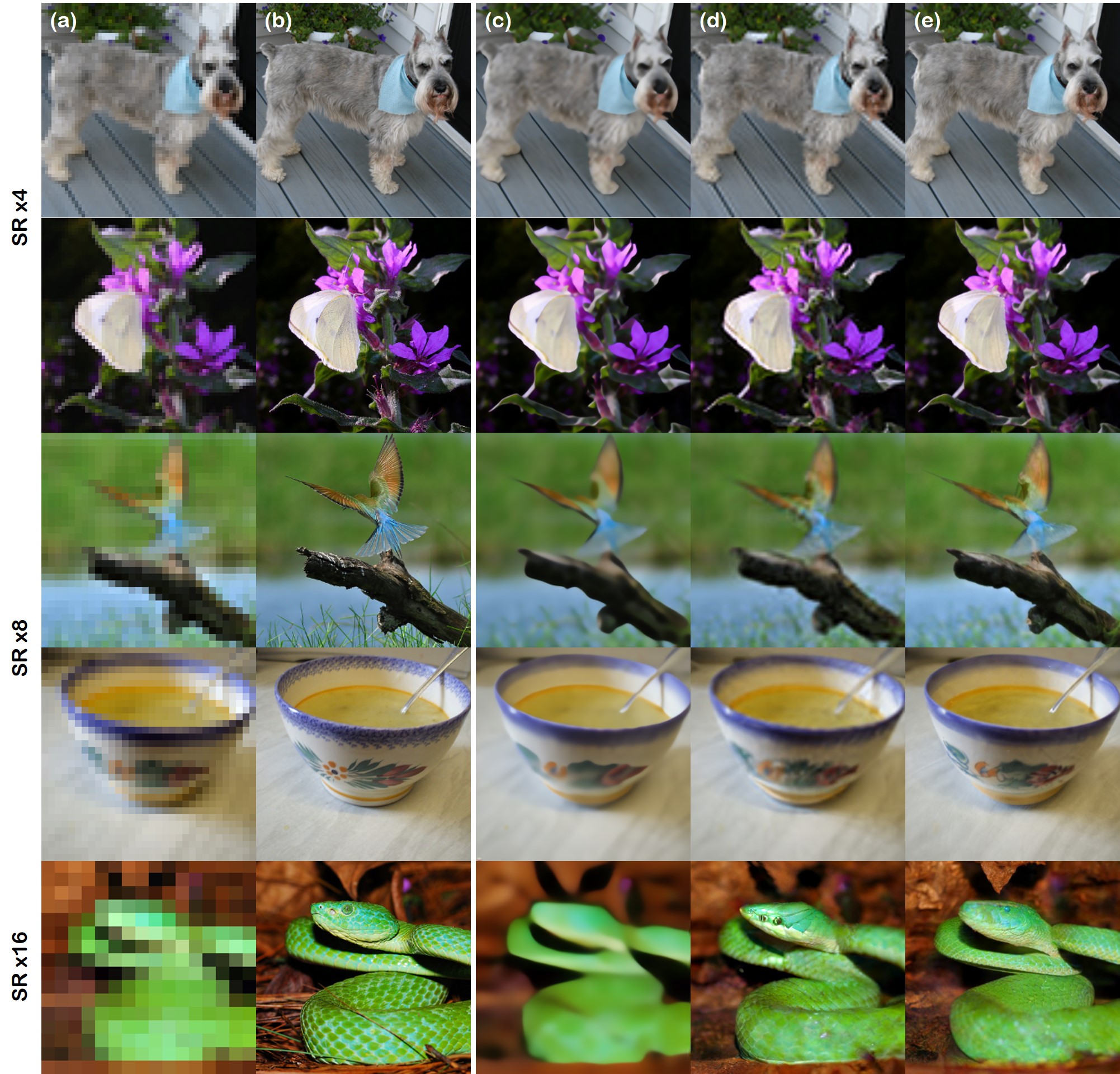}
    \caption{{Results of SR task on ImageNet256 validation dataset. (a) LR image, (b) ground truth, (c) ESRGAN, (d) ILVR (1000 steps), (e) CCDF (100, 200, 300 steps for $\times$4, $\times$8, $\times$16 SR, respectively.)}}
	\label{fig:SR_results_imagenet}
\end{figure*}

\noindent
\textbf{{Experiments with ImageNet.}}~{ImageNet\cite{deng2009imagenet} contains diverse categories of natural images, and are known to be much harder to model, due to its highly multimodal nature. We try to examine if CCDF scales even to this challenging task, using a pre-trained model provided in the guided-diffusion github repository\footnote{\href{https://github.com/openai/guided-diffusion}{https://github.com/openai/guided-diffusion}}. As with other experiments with FFHQ or AFHQ dataset, we train an ESRGAN model for each SR factor, and use it as our initialization strategy. In Fig.~\ref{fig:SR_results_imagenet}, we can see that our CCDF strategy outperforms ILVR using full reverse diffusion, and also vastly improves the image quality of ESRGAN, which is our initialization.}

\subsection{Inpainting}

\begin{table}[!hbt]
    \centering
    \resizebox{0.5\textwidth}{!}{
    \begin{tabular}{c|cccccc}
    \hline
    $t_0$ & 0.05 & 0.1 & 0.2 & 0.5 & 0.75 & 1.0~\cite{song2020score} \\ \hline\hline
    Box 96   & 46.03 & \textbf{45.93} & \underline{45.99} & 46.14 & 48.05 & 48.61 \\ 
    Box 128  & 50.41 & \underline{50.05} & \textbf{49.77} & 51.65 & 54.49 & 59.27 \\
    Box 160  & 61.77 & \underline{59.62} & \textbf{57.99} & 61.04 & 67.50 & 78.50 \\ \hline
    \end{tabular}
    }
    \caption{FID($\downarrow$) scores on FFHQ test set for inpainting task with varying $t_0$ values. $t_0 = 1.0$ is the baseline method without any acceleration used in~\cite{song2020score}. Numbers in boldface, and underline indicate the best, and the second best scores.}
    \label{tab:FID_inpaint_FFHQ}
\end{table}

\noindent
\textbf{{Dependence on $t_0$.}}~{As in Table~\ref{tab:FID_SR_FFHQ}, we compare the FID score of reconstructions for the inpainting task, as we vary the $t_0$ values in table~\ref{tab:FID_inpaint_FFHQ}. We notice similar results from the SR task, in the sense that there always exist $t_0 \in (0, T)$ which gives higher scores than using full diffusion. With relatively small boxes, we see that $t_0 = 0.1$ is optimal, whereas we typically need more diffusion steps for larger boxes.}

\begin{figure}[!hbt]
    \centering\includegraphics[width=8cm]{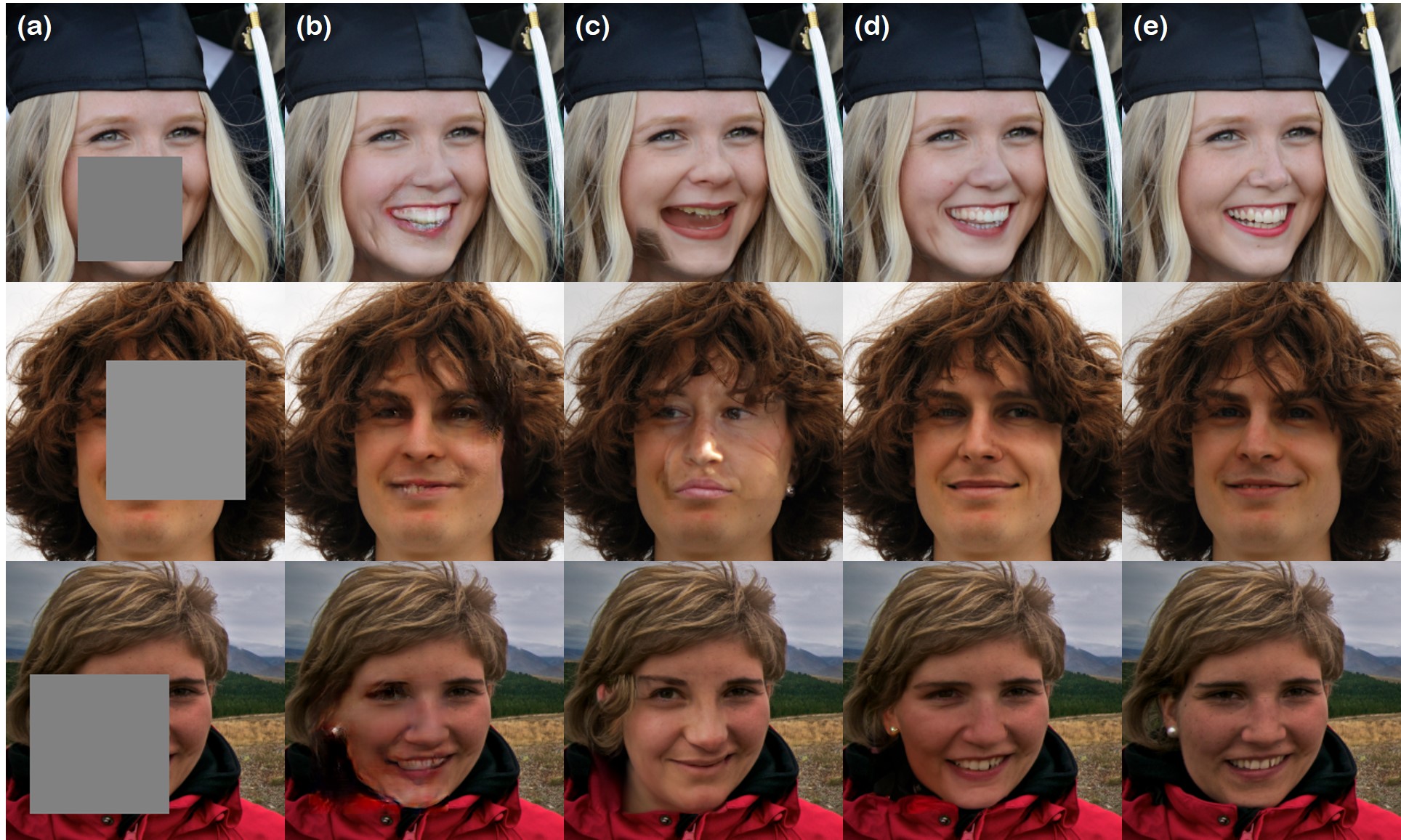}
    \caption{Comparison on inpainting task: (a) Input image, (b) SN-PatchGAN~\cite{yu2019free}, (c) score-SDE~\cite{song2020score} using 1000 steps from $T=1$, (d) CCDF using 200 steps from $t_0=0.2$, (e) Ground Truth.}
	\label{fig:inpaint_comparison}
\end{figure}

\begin{figure}[!h]
    \centering\includegraphics[width=8cm]{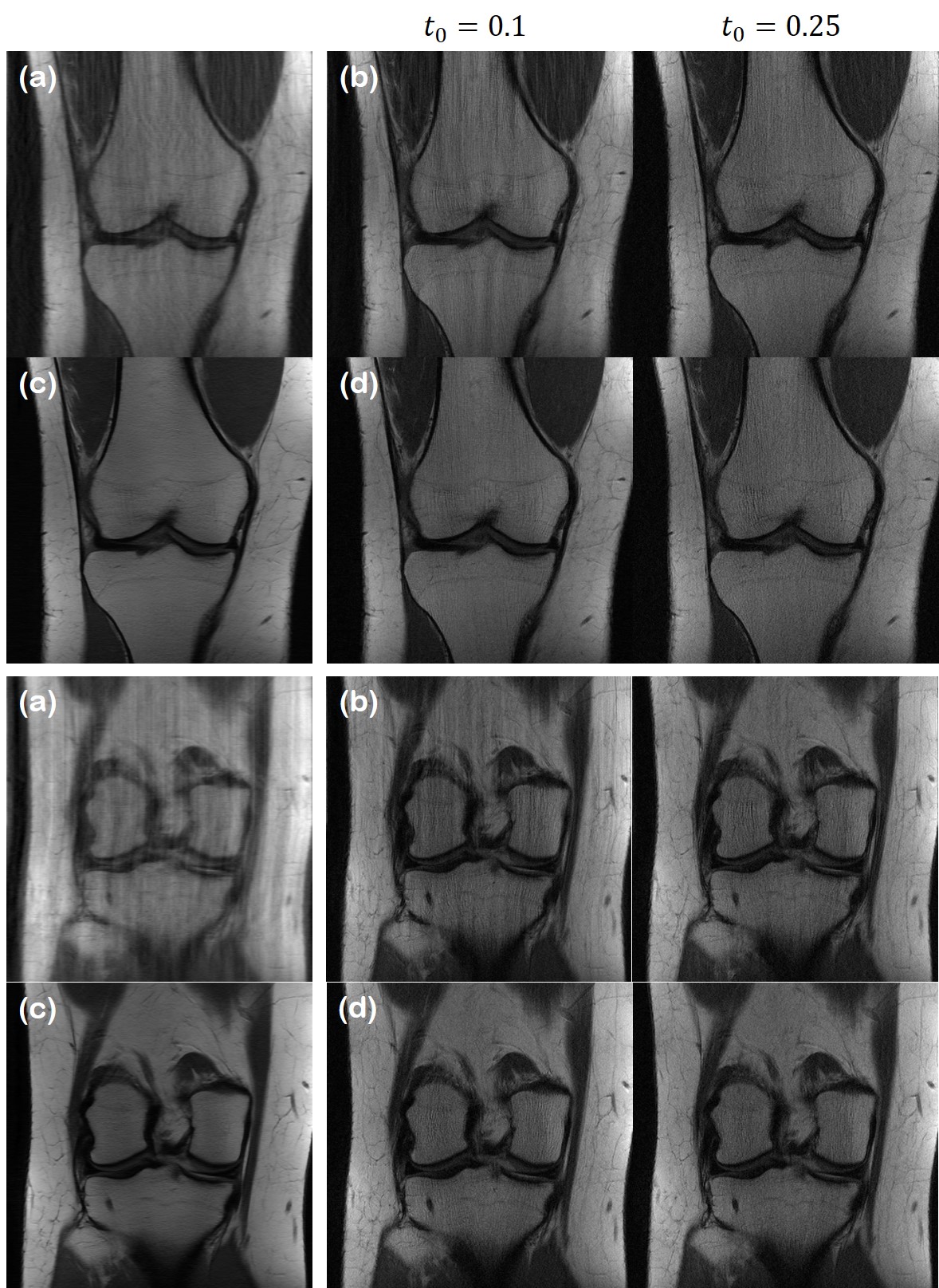}
    \caption{Ablation study on using different initializations for forward diffusion ($\times 6$ 1D Gaussian sampling). (a) Vanilla initialization, (b) corresponding results with the proposed method. (c) NN initialization, (d) corresponding results with the proposed method.}
	\label{fig:MR_ablation_init}
\end{figure}

\noindent
\textbf{{Comparison study.}}~{We compare the proposed CCDF strategy with SN-PatchGAN~\cite{yu2019free}, and score-SDE~\cite{song2020score} using 1000 steps in Fig.~\ref{fig:inpaint_comparison}. For SN-PatchGAN in Fig.~\ref{fig:inpaint_comparison} (b), we often see highly unrealistic details e.g. near the mouth. Note that SN-patchGAN serves as the initialization point for CCDF in inpainting. Leveraging this imperfect initialization, the proposed method is able to provide reconstructions that are highly realistic, as can be seen in fig.~\ref{fig:inpaint_comparison} (d). It is also notable that score-SDE using full diffusion more often than not produces results that are incoherent with the known regions (see second row of Fig.~\ref{fig:inpaint_comparison} (c)), while the proposed method stably outputs coherent results.}

\begin{table}[!hbt]
    \centering
    \resizebox{0.4\textwidth}{!}{
    \begin{tabular}{c|c|c|cccc}
    \hline
         & SR & \thead{$t_0=$} & 0.1 & 0.2 & 0.5 & 1.0 \\ \hline\hline
    \multirow{6}{*}{FFHQ} & \multirow{2}{*}{$\times 4$} & vanilla & 78.39 & 66.77 & 64.25 & 63.14 \\
    & & NN init. & \textbf{60.90} & \textbf{60.91} & 64.04 & 63.31 \\ \cline{2-7}
    & \multirow{2}{*}{$\times 8$} & vanilla & 116.42 & 93.06 & 82.39 & 78.91 \\
    & & \thead{NN init.} & 78.13 & \textbf{75.76} & 79.34 & \underline{77.34} \\ \cline{2-7}
    & \multirow{2}{*}{$\times 16$} & vanilla & 184.70 & 135.20 & 96.15 & 92.32 \\
    & & \thead{NN init.} & 101.79 & 92.59 & \textbf{88.09} & \underline{88.49} \\ \hline
    \multirow{6}{*}{AFHQ} & \multirow{2}{*}{$\times 4$} & vanilla & 19.14 & 18.66 & 18.08 & 18.70 \\
    & & NN init. & \textbf{15.53} & \underline{17.14} & 19.06 & 18.10 \\ \cline{2-7}
    & \multirow{2}{*}{$\times 8$} & vanilla & 48.87 & 39.88 & 33.28 & 34.84 \\
    & & NN init. & \underline{33.47} & \textbf{32.30} & 33.65 & 33.50 \\ \cline{2-7}
    & \multirow{2}{*}{$\times 16$} & vanilla & 96.01 & 72.22 & 47.42 & 47.28 \\
    & & NN init. & 63.27 & 51.13 & \textbf{44.18} & \underline{45.17} \\ \hline
    \end{tabular}
    }
    \caption{{FID($\downarrow$) scores for SR tasks with different initialization strategies.}}
    \label{tab:FID_SR_initialization}
\end{table}

\noindent
\textbf{{Ablation study.}}~We perform an ablation study comparing the effect of different initialization strategy. Table~\ref{tab:FID_SR_initialization} shows the difference in the results when using vanilla initialization with {the corrupted image}, and NN initialization. We see that with all $t_0$, NN initialization performs marginally better than vanilla initialization. The difference becomes clearer as we decrease the value of $t_0$ to 0.1. The same ablation study was performed also for MRI reconstruction task, and is illustrated in Figure~\ref{fig:MR_ablation_init}. We see similar trend as in the SR task.

\begin{figure*}[!h]
    \centering\includegraphics[width=17cm]{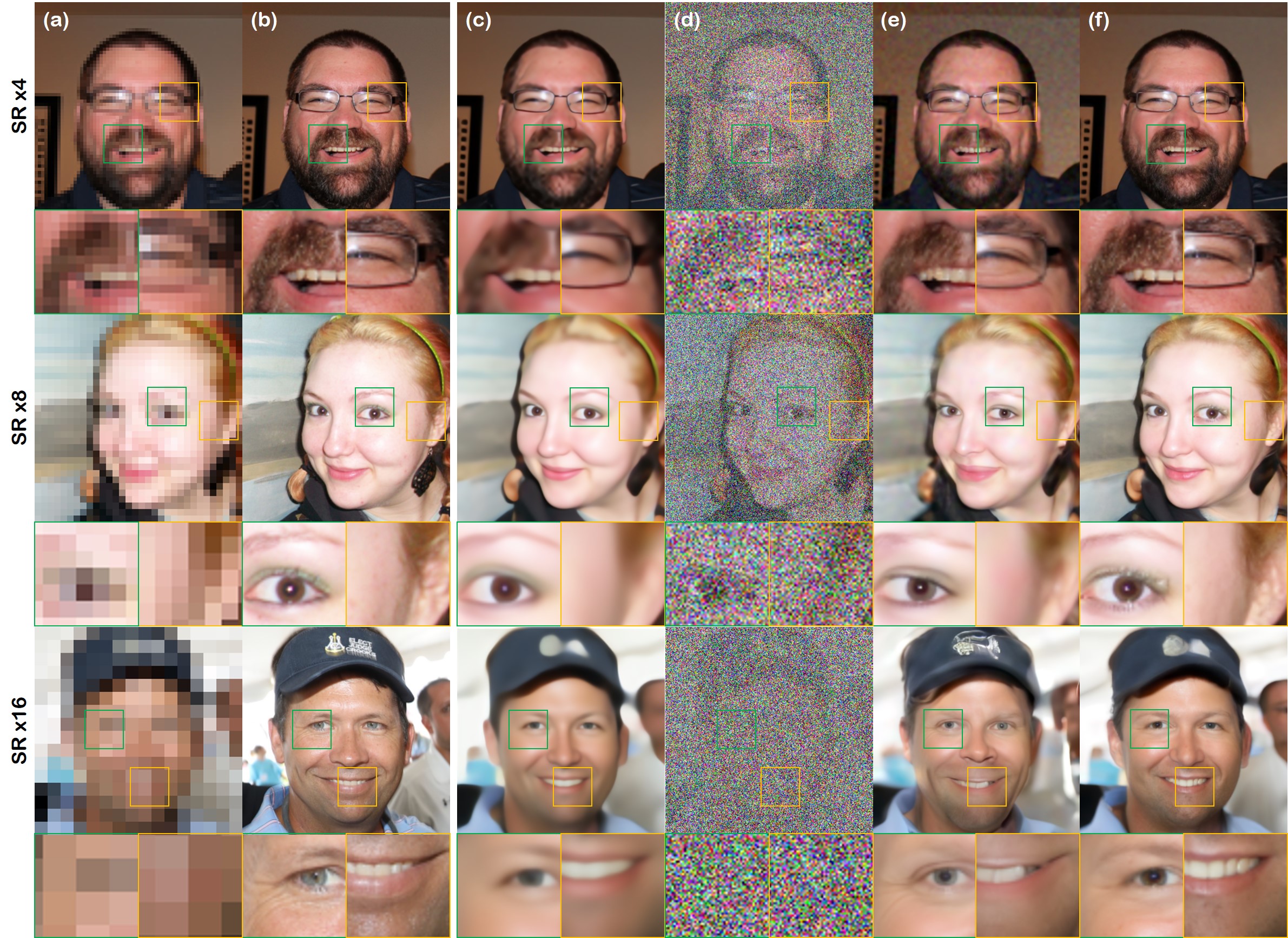}
    \caption{{Results of super-resolution on FFHQ 256$\times$256 data. The first, second and third row denote $\times 4$ SR, $\times 8$ SR, and $\times 16$ SR, respectively. (a) LR input, (b) Ground Truth, (c) ESRGAN~\cite{wang2018esrgan}, (d) SR3~\cite{saharia2021image} with 20 diffusion steps ($N = 20, \Delta t = 0.05$), (e) ILVR~\cite{choi2021ilvr} with 20 diffusion steps($N = 20, \Delta t = 0.05$), (f) proposed method (CCDF) with 20 diffusion steps ($N = 100, t_0 = 0.2$).}}
	\label{fig:SR_results_ffhq_20}
\end{figure*}

\begin{figure*}[!h]
    \centering\includegraphics[width=17cm]{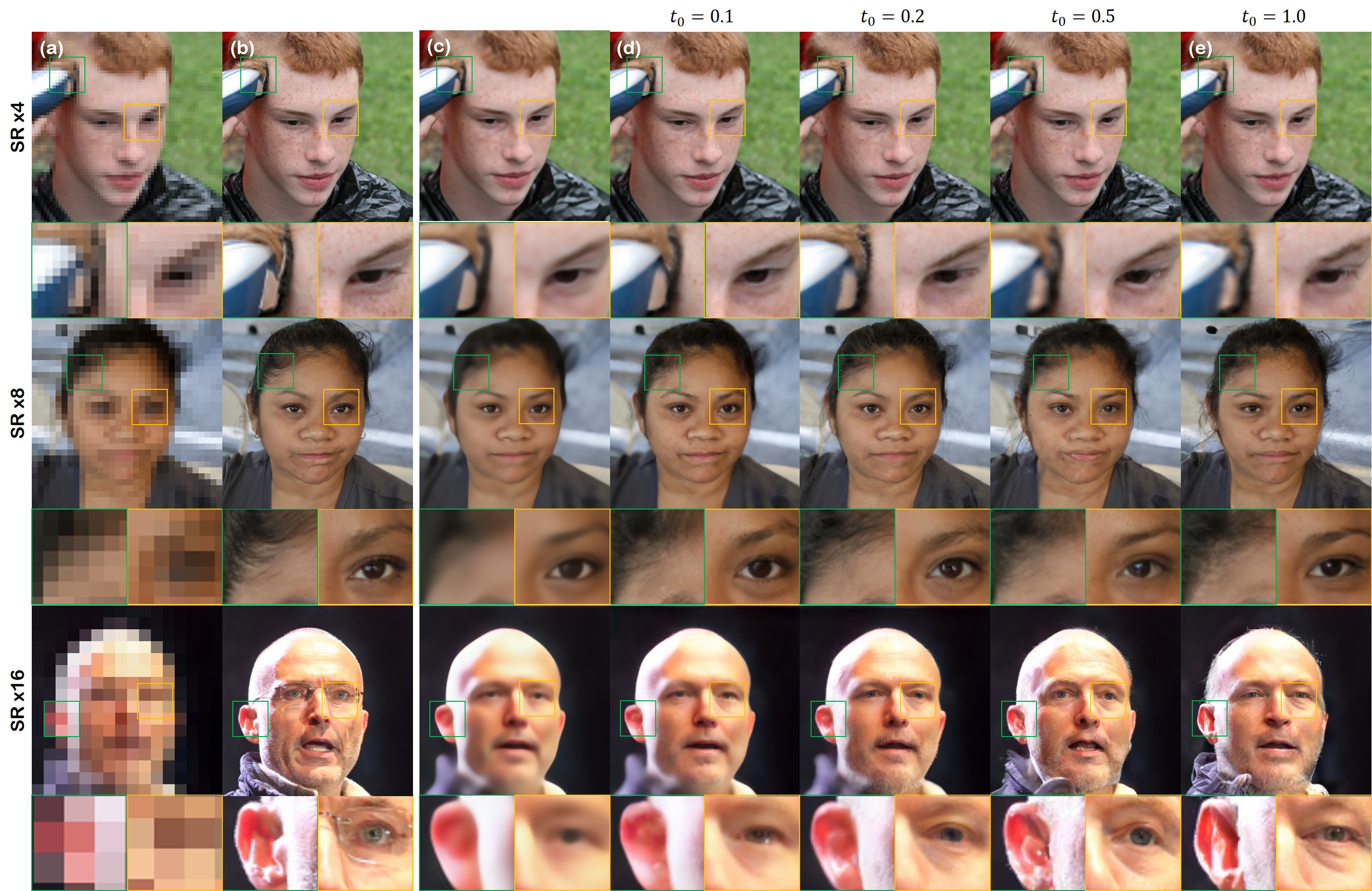}
    \caption{Results of super-resolution on FFHQ 256$\times$256 data. The first, second and third row denote $\times 4$ SR, $\times 8$ SR, and $\times 16$ SR, respectively. (a) LR input, (b) Reference, (c) ESRGAN~\cite{wang2018esrgan}, (d) proposed method (CCDF) with varying $t_0$ values, and (e) ILVR ($t_0 = 1.0$)~\cite{choi2021ilvr}.}
	\label{fig:SR_results_ffhq2}
\end{figure*}

\begin{figure*}[!hbt]
    \centering\includegraphics[width=17cm]{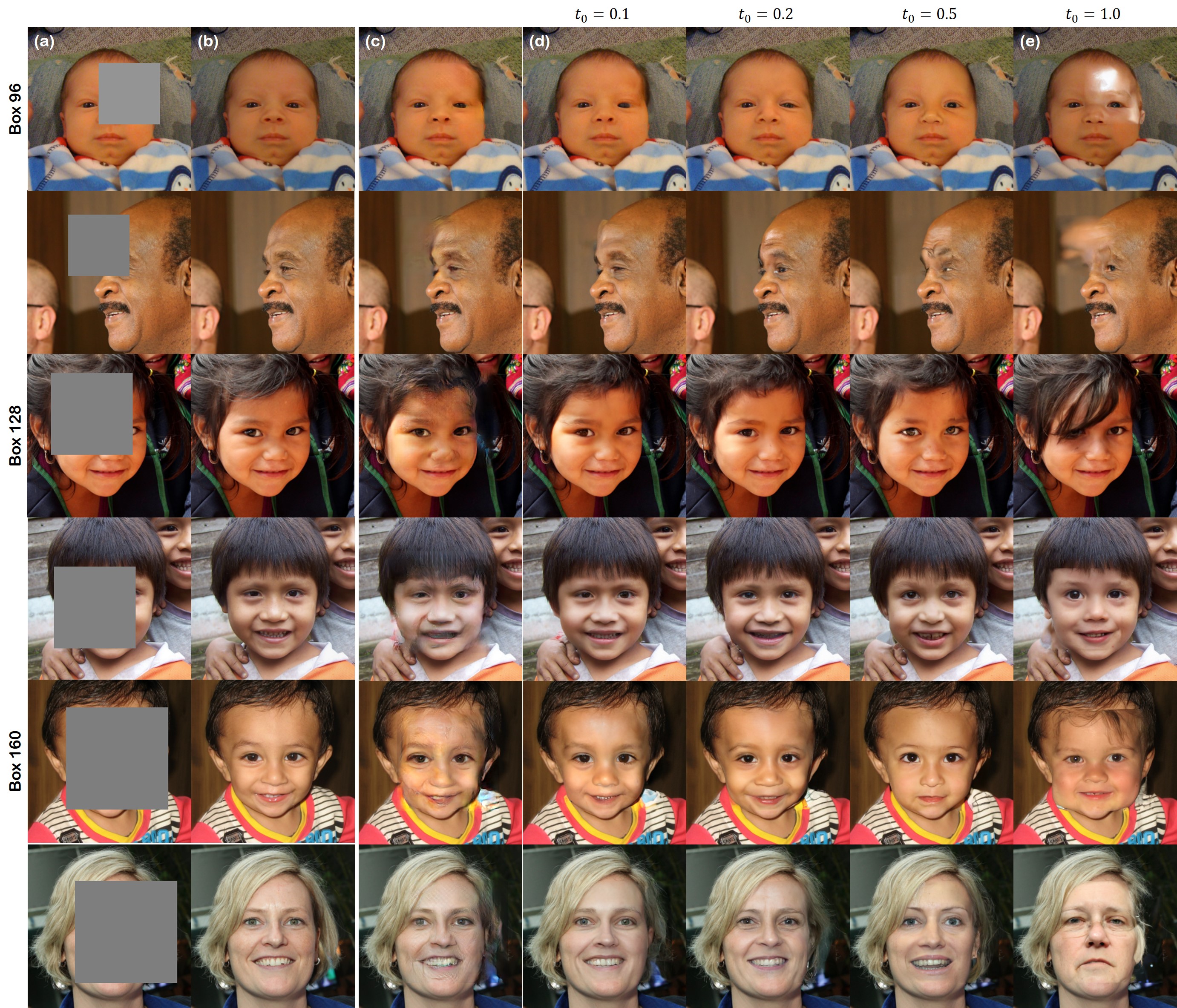}
    \caption{Additional results of inpainting on FFHQ 256$\times$256 data. The first, second and third row denote masks of size $96\times96, 128\times128$ and $160\times160$, respectively. (a) Masked image, (b) SN-patchGAN~\cite{yu2019free}, (c) proposed method (CCDF) with varying $t_0$ values, and (d) Score-SDE ($t_0 = 1.0$)~\cite{song2020score}}
	\label{fig:inpaint_results_ffhq2}
\end{figure*}

Furthermore, we provide additional qualitative results of each task on various datasets, focusing mainly on showing the trend of reconstruction results as we vary the value of $t_0$. {In Figure~\ref{fig:SR_results_ffhq_20}, we compare the achievable image quality by fixing the number of reverse diffusion steps to 20. Consistent with what we saw in Figure~\ref{fig:SR_results}, we see that our method largely outperforms the other diffusion model-based methods.}
In Figure~\ref{fig:SR_results_ffhq2}, Figure~\ref{fig:inpaint_results_ffhq2} respectively, we see that we can stably arrive at a feasible solution with different values of $t_0 \in [0.1, 0.5]$, typically requiring higher values of $t_0$ for severer degradation.

\section{{Validity of assumption}}
\label{sec:supp_v}

{In Lemma~\ref{lem:dsm}, we assumed that $s_\theta(\xb_i, t) = \frac{\partial}{\partial \xb_i}\log p_{0i}(\xb_i|\xb_0)$. In this section, we briefly show that the assumption is valid. In the theoretic side,~\cite{alain2014regularized} showed that given an {\em optimal} reconstruction function $r^*_\sigma(\xb_t)$ trained with denoising autoencoder loss asymptotically behaves as}
\begin{align*}
    r^*_\sigma(\xb_t) &= \xb_t + \sigma^2\frac{\partial\log p(\xb_t)}{\partial\xb_t} + o(\sigma^2),\quad \sigma \rightarrow 0\\
    s^*_\theta(\xb_t) &= \frac{r^*_\sigma(\xb_t) - \xb_t}{\sigma^2} = \frac{\partial \log p(\xb_t)}{\partial\xb_t} + o(1),\quad \sigma \rightarrow 0,
\end{align*}
{where the equation emphasizes the behavior of the optimal {\em score} function, regarding how it was parameterized. This asymptotic behavior hints that the error will be small, especially when $\sigma$ approaches zero. In our case, $\sigma \rightarrow 0$ as $t \rightarrow 0$, and hence the behavior holds near $t = 0$.}

\begin{figure}[!h]
    \centering\includegraphics[width=8cm]{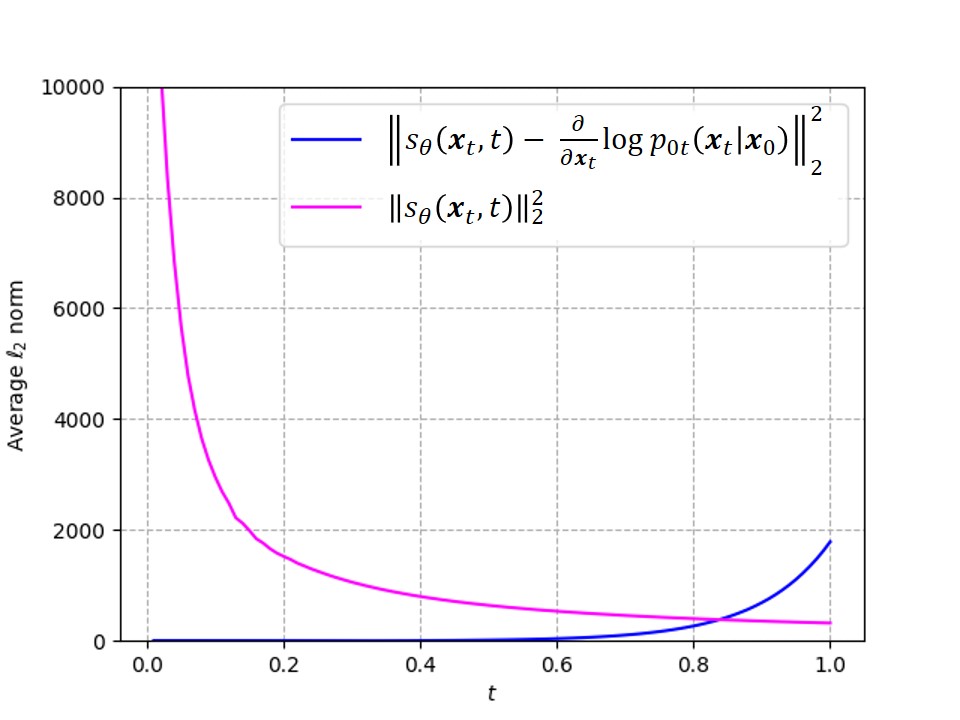}
    \caption{{Average value of $\|s_\theta(\xb_t, t) - \frac{\partial}{\partial \xb_t} \log p_{0t}(\xb_t|\xb_0)\|_2^2$, and the norm of $\|s_\theta(\xb_t, t)\|_2^2$ at time $t$, on fastMRI 320$\times$320 test set (1K samples).}}
	\label{fig:validity}
\end{figure}

{In order to numerically validate such error, we conducted an experiment to calculate the actual average error norm, illustrated in Fig.~\ref{fig:validity}. Here, we see that the error norm mostly stays at very low values across the range. We do observe that the magnitude of error inevitably increases when the noise level is too large, so the error grows where $t > 0.5$. Nevertheless, we note that most of our contraction analysis stays in the $t \in [\epsilon, 0.5]$ regime, and hence the assumption made in Lemma A.1. is practical.}

\end{document}